\documentclass[preprint,letter]{revtex4}              

\usepackage{amsmath}
\usepackage{graphicx,epsfig}
\usepackage[T1]{fontenc}
\usepackage[latin1]{inputenc}
\usepackage[hypertex]{hyperref}
\usepackage[left=1in,right=1in,letterpaper,centering]{geometry}
\usepackage{setspace}
\hypersetup{colorlinks,bookmarksopen,bookmarksnumbered,citecolor=blue}

\usepackage{booktabs}

\begin{document}
\def\deg{^\circ}
\def\half{ \small {1 \over 2}}
\bibliographystyle{naturemag}



\title{X-ray Near Field Speckle: Implementation and Critical Analysis}

\author {Xinhui Lu}
\affiliation{Department of Physics, Yale University, New Haven, Connecticut 06511}
\affiliation{Present address: Department of Condensed Matter Physics and Materials Science, Brookhaven National Laboratory, Upton, New York 11973}
\author{S. G. J. Mochrie}
\affiliation{Department of Physics, Yale University, New Haven,
Connecticut 06511\\Department of Applied Physics, Yale
University, New Haven, Connecticut 06511}
\author{S. Narayanan}
\affiliation{Advanced Photon Source, Argonne National Laboratory,
Argonne, IL 60439}
\author{A. R. Sandy}
\affiliation{Advanced Photon Source, Argonne National Laboratory,
Argonne, IL 60439}
\author{M.  Sprung}
\affiliation{Advanced Photon Source, Argonne National Laboratory,
Argonne, IL 60439}
\affiliation{Present address: Present address: HASYLAB at DESY, D-22607 Hamburg, Germany.}

\begin{abstract}
We have implemented the newly-introduced, coherence-based technique of x-ray near-field speckle (XNFS) at 8-ID-I at the Advanced Photon Source. In the near field regime of high-brilliance synchrotron x-rays scattered from a sample of interest, it turns out, that, when the scattered radiation and the main beam both impinge upon an x-ray area detector, the measured intensity shows low-contrast speckles, resulting from interference between the incident and scattered beams. We built a micrometer-resolution XNFS detector with a high numerical aperture microscope objective and demonstrate its capability for studying static structures and dynamics at longer length scales than traditional far field x-ray scattering techniques. Specifically, we characterized the structure and dynamics of dilute silica and polystyrene colloidal samples. Our study reveals certain limitations of the XNFS technique, which we discuss.
\end{abstract}

\maketitle

\section{Introduction}

Although Small Angle X-ray Scattering (SAXS) and X-ray Photon Correlation Spectroscopy (XPCS) have succeeded in exploring the structure and dynamics of many interesting systems, the length scale of the observable systems is generally limited to a range from several nm to 100 nm, corresponding to a wavevector range of $10^{-3}$ \AA $^{-1}$ to 0.1 \AA$^{-1}$ (an angular range of 0.1$^\circ$ to 10$^\circ$)~\cite{dierker:95,mochrie:97,PontoniPRL2003,falus:04,PhysRevLett.97.075505,XluPRL08}. Special difficulties are encountered when exploring the lower limit of the angular range, since to isolate the weak scattering from the strong direct beam, it is necessary to block the direct beam and extraneous scattering from slits, etc., which sets a boundary for the smallest detected angle. In order to probe longer length scales with x-rays, typically a Bonse-Hart camera is used, which can access a wavevector ($q$) range of $10^{-4}$ \AA $^{-1}$ to 0.1 \AA$^{-1}$~\cite{Diat1995566,ilavsky:1660,Narayanan20011005}. Conventional Bonse-Hart camera is one-dimensional collimated which is not suitable for anisotropic samples. At a cost of reduced scattering intensity, there are several papers describing a 2D-collimated Bonse-Hart camera~\cite{BonseZP1966,Konishi:xs0086,Ilavsky:ce5052}. They demand a de-convolution procedure and are inefficient in comparison to area-detector based method available for larger wavevectors. And the scanning procedure makes it difficult for time-resolved measurements. In addition, previous works \cite{doi:10.1021/la001184y,Shinohara:sy6016} show that ultra small-angle could be achieved by using a very long sample-to-detector distance or a very small beam stop. In this case, additional interpolations with SAXS data are required because of the limitation of the field of view. The recently-introduced, coherence-based technique of X-ray Near-Field Speckle (XNFS) technique potentially offers an improved means of characteristically large length scale structure with x-rays~\cite{cerbino:natphys}. In addition, XNFS is able to extend x-ray measurements to wavevectors (length scales) at least an order of magnitude smaller (larger) than may be achieved by a Bonse-Hart camera.

The principle of XNFS is as follows. When a coherent or partially coherent radiation impinges on a disordered material consisting of a number of scatterers at random locations, a random set of phase shifts will be induced on the scattering beam. As a consequence, a grainy pattern will be observed in the scattered beam a certain distance away from the material. This pattern is called a speckle pattern~\cite{sutton:91}. In the near field region, under conditions where the scattered radiation and the transmitted beam simultaneously impinge upon an x-ray area detector, high-quality speckles can be also observed result from coherent interference between the incident and scattered beams. These speckles are called x-ray near-field speckle (XNFS) in analogy to the NFS that was initially exploited using laser sources~\cite{PhysRevLett.85.1416}.

If instead of laser, one uses a high-brilliance x-ray source, it then becomes possible to study dense, optically turbid and/or absorbing media, in a range of length scales where no other x-ray or optical methods are applicable. To date, there exists a single manuscript describing the extension of NFS into the x-ray regime by Cerbino {\em et al}~\cite{cerbino:natphys}. They showed, the spatial power spectrum of x-ray NFS is in principle simply and directly related to the sample's structure factor [$S(q)$] in the range of wavevectors from $10^{-5}$ \AA$^{-1}$ or less, to $10^{-3}$ \AA$^{-1}$ or larger. Equivalently, x-ray NFS measures the density-density correlation function [$g_1(r)$] from length scales of $6 \times 10^4$ \AA~or more, to $1 \times 10^3 $ \AA~or less. In addition, the evolution of the heterodyne speckle pattern in time determines the sample's intermediate scattering function [$S(q, t)$], and its spatial Fourier transform, $g_1(r, t)$. Here, we will demonstrate the implementation of XNFS measurements at beamline 8-ID at the Advanced Photon Source (APS) at Argonne National Laboratory, and will explore and discuss the utility and drawbacks of this method for studies of colloidal suspensions.

\section{Basic Theory}

In this section, we present a derivation of what we may expect to measure in XNFS experiments. We envision a sample with density $\rho(x, y, z,t)$ and a detector located in the plane $z = z'$, so that $x'$ and $y'$ specify a given detector pixel. Then, we may write for the amplitude scattered by volume element $dxdydz$ at $(x, y, z)$ to the detector pixel at $(x',y',z')$ at time $t$:
\[d{a_s} = i{r_0}{a_0}{e^{ikz - ik\int_0^z {\delta (x,y,s)ds}  - k\int_0^z {\beta (x,y,s)ds} }}\rho (x,y,z,t)dxdydz\frac{{{e^{ikR}}}}
{R}\]
where $r_0$ is the Thomson radius, ${a_0}{e^{ikz - ik\int_0^z {\delta (x,y,s)ds}  - k\int_0^z {\beta (x,y,s)ds} }}$ is the amplitude of the incident wave at z, and
\[R = \sqrt {{{(x' - x)}^2} + {{(y' - y)}^2} + {{(z' - z)}^2}} \]
$\delta$ and $\beta$ are the real and imaginary parts of the x-ray refractive index.

For a sufficiently uniform sample, for which the $z$-integrals of $\delta$ and $\beta$ are independent of $x$ and $y$, and ignoring the phase part, we may write
\begin{equation}\label{eq:scattamp}
    d{a_s} =i{r_0}{a_0}{e^{ikz}}\rho(x,y,z,t)dxdydz\frac{{{e^{ikR}}}} {R}{e^{ - z/\Lambda }}
\end{equation}
where $\Lambda$ is the x-ray absorption length. Eq.\ref{eq:scattamp} represents a quite different regime than that used to interpret x-ray imaging experiments. Such imaging experiments instead rely on the $x$ and $y$ dependence of $\int_0^z {\delta (x,y,s)ds}$ and $\int_0^z {\beta (x,y,s)ds}$ to create an image of the sample. Henceforth, we
will neglect explicit mention of absorption, but otherwise take Eq. \ref{eq:scattamp}, as our expression for the scattered amplitude.

In the near-field regime, in which scattered x-rays interfere with the incident beam within the coherence area of the incident beam, we are necessarily concerned with values of $x'-x$ and $y'-y$ that are on the order of 100 $\rm{\mu} m$ or less, and values of $z'-z$ that are on the order of several millimeters or more, so that $z'-z \gg x'-x$ and $z'-z \gg y'-y$. It follows that
\begin{equation}
    R \simeq (z' - z) + \frac{{{{(x' - x)}^2} + {{(y' - y)}^2}}}
{{2(z' - z)}}
\end{equation}
and, therefore,
\begin{equation}
    d{a_s} \simeq i{r_0}{a_0}\rho (x,y,z,t)dxdydz{e^{ikz'}}\frac{{{e^{ik{{\left[ {{{(x' - x)}^2} + {{(y' - y)}^2}} \right]} \mathord{\left/
 {\vphantom {{\left[ {{{(x' - x)}^2} + {{(y' - y)}^2}} \right]} {\left[ {2(z' - z)} \right]}}} \right.
 \kern-\nulldelimiterspace} {\left[ {2(z' - z)} \right]}}}}}}
{{z' - z}}
\end{equation}
To determine the total amplitude scattered to ($x',y',z'$), it is simply necessary to integrate over the volume ($V$) of the sample, i.e.
\begin{equation}
    {a_s} = i{r_0}{a_0}\int_V {dxdydz\rho (x,y,z,t){e^{ikz'}}\frac{{{e^{ik{{\left[ {{{(x' - x)}^2} + {{(y' - y)}^2}} \right]} \mathord{\left/
 {\vphantom {{\left[ {{{(x' - x)}^2} + {{(y' - y)}^2}} \right]} {\left[ {2(z' - z)} \right]}}} \right.
 \kern-\nulldelimiterspace} {\left[ {2(z' - z)} \right]}}}}}}
{{z' - z}}}
\end{equation}

Heterodyne near-field speckle involves interference between the scattered beam and the incident beam of amplitude $a_{z'}$ . Thus, the intensity at time $t$ recorded at ($x',y',z'$) is
\begin{equation}
I(x',y',t) = {\left| {{a_{z'}}} \right|^2} + a_{z'}^ * {a_s} + {a_{z'}}a_s^ *  + {\left| {{a_s}} \right|^2} \simeq {\left| {{a_{z'}}} \right|^2} + a_{z'}^ * {a_s} + {a_{z'}}a_s^ *
\end{equation}
where at the detector $a_{z'} = a_0e^{ikz'}$ and we have taken $a_s \ll a_{z'}$.

Therefore, the measured intensity is
\begin{align}\label{eq:int1}
  &I(x',y',t) =  {\left| a_0 \right|^2}\left(\right.  1+  i{r_0}\int_V dxdydz \\ \nonumber
  &\left[\rho (x,y,z,t) \tfrac{e^{{ik\left[ {{{\left( {x' - x} \right)}^2} + {{\left( {y' - y} \right)}^2}} \right]} /{\left[ {2\left( {z' - z} \right)} \right]}}}
{z' - z}  \right.\hfill  \\ \nonumber
   &\left.-  {\rho ^ * }(x,y,z,t) \tfrac{e^{ - {{ik\left[ {{{\left( {x' - x} \right)}^2} + {{\left( {y' - y} \right)}^2}} \right]} / {\left[ {2\left( {z' - z} \right)} \right]}}}}
{z' - z} \left.\right)\right] \hfill \\ \nonumber
   &= {\left| {{a_0}} \right|^2}\left(1 - 2{r_0}\int_V dxdydz \right. \\ \nonumber
   &\left[\rho '(x,y,z,t) \tfrac{\sin \left( {{k\left[ {{{\left( {x' - x} \right)}^2} + {{\left( {y' - y} \right)}^2}} \right]} / {\left[ {2\left( {z' - z} \right)} \right]}} \right)}{z' - z} \right.    \hfill \\ \nonumber
   & \left.+ \rho ''(x,y,z,t) \tfrac{\cos \left( {{k\left[ {{{\left( {x' - x} \right)}^2} + {{\left( {y' - y} \right)}^2}} \right]} /{\left[ {2\left( {z' - z} \right)} \right]}} \right)}{z' - z}\right] \left.\right) \hfill
\end{align}
where $\rho'$ and $\rho''$ are the real and imaginary (absorptive) parts of the electron density, respectively.

Eq.~\ref{eq:int1} implicitly assumes perfect transverse coherence. To incorporate the effect of a finite
transverse coherence length, it is necessary to introduce a mutual coherence function:
\begin{equation}
{\gamma _{12}}\left( {x' - x,y' - y} \right) \simeq {e^{ - \left[ {{{(x' - x)}^2}/(2{\xi_x ^2}) + {{(y' - y)}^2}/(2{\xi_y ^2})} \right]}}
\end{equation}
where $\xi_x$ and $\xi_y$ are the transverse coherence lengths in the x- and y-directions, respectively. Incorporating
the effect of partial coherence, Eq.~\ref{eq:int1} becomes

\begin{align}\label{eq:I(x)}
I(x',y',t)
   &\simeq {\left| {{a_0}} \right|^2}  \left(1 - 2{r_0}\int_V dxdydz{e^{ - \left[ {\tfrac{{(x' - x)}^2}{2{\xi_x ^2}} + \tfrac{{(y' - y)}^2}{2{\xi_y ^2}}} \right]}}  \right.\\ \nonumber
   &\left(\rho'(x,y,z,t)  {\tfrac{{\sin \left( {k\left[ {{{\left( {x' - x} \right)}^2} + {{\left( {y' - y} \right)}^2}} \right]} \left/
 { { 2\left( {z' - z} \right)} } \right.\right)}}{{z' - z}}}  \right.\\ \nonumber
  &+ \left.\left.\rho''(x,y,z,t){\tfrac{{\cos \left( {k\left[ {{{\left( {x' - x} \right)}^2} + {{\left( {y' - y} \right)}^2}} \right]} \left/
 { { 2\left( {z' - z} \right)} } \right.\right)}}{{z' - z}}}  \right)\right)
\end{align}

The first term of Eq. \ref{eq:I(x)} is constant and the second term of Eq. \ref{eq:I(x)} is a convolution in terms of real space variables. Therefore, in Fourier space, the first term becomes a $\delta$-function at the origin, while the second term becomes a product. Therefore, in terms of the Fourier transform variables, $q$ and $p$ ($q\neq0 $ and $p\neq 0$), in the realistic case that the $z$-variations in the sample density occur on length scales less than $k/(q^2)$, which is typically  hundreds of micrometers or more, it may further be shown that the Fourier transformed intensity is given by
\begin{align}\label{eq:I(q)}
    & \tilde I(q,p,t) \simeq {\left| {{a_0}} \right|^2}2{r_0} \\ \nonumber
    & \left( \int_\Lambda  dz {e^{\frac{{ - {{(z' - z)}^2}}}
{2}\left( {\frac{{{q^2}{\xi_x ^2}}}
{{{{(z' - z)}^2} + {k^2}{\xi_x ^4}}} + \frac{{{p^2}{\xi_y ^2}}}
{{{{(z' - z)}^2} + {k^2}{\xi_y ^4}}}} \right)}}\tfrac{{\xi_x \xi_y }}
{{{{({{(z' - z)}^2} + {k^2}{\xi_x ^4})}^{1/4}}{{({{(z' - z)}^2} + {k^2}{\xi_y ^4})}^{1/4}}}}   \right.\\ \nonumber
& \left(  \rho '(q,p,z,t)    \sin \left[  \Phi \right]  +  \rho ''(q,p,z,t)   \left. \cos \left[ \Phi \right] \right) \right)
\end{align}
where $\Lambda$ is the thickness of the sample, $\rho(q,p,z,t)$ is mixed in real and reciprocal space and the phase factor $\Phi$ is equal to $ {\tfrac{1}
{2}\left( { - \tfrac{{k{q^2}(z' - z){\xi_x ^4}}}
{{{{(z' - z)}^2} + {k^2}{\xi_x ^4}}} - \tfrac{{k{p^2}(z' - z){\xi_y ^4}}}
{{{{(z' - z)}^2} + {k^2}{\xi_y ^4}}} + {{\tan }^{ - 1}}\tfrac{{(z' - z)}}
{{k{\xi_x ^2}}}+{{\tan }^{ - 1}}\tfrac{{(z' - z)}}
{{k{\xi_y ^2}}}} \right)}$. Notice that the $\tan ^{-1}$ terms in the phase factor describe an on-axis phase jump of a focused beam (Gouy effect)~\cite{Gouy1890} and underline the well known fact that field correlations propagate as the radiation field does. So these terms account for the change in the phase in the interference between scattered beam and transmitted beam.

Eq.~\ref{eq:I(q)}, which stands as a one-dimensional convolution, may be further simplified by the following argument: the $z$-variations in $\rho(q,p,z,t)$ occur on length scales set by the sample's structure, namely on the order of tens of micrometers or less. On the other hand, the $z$-variations in the remainder of the integrand occur on a length scale given by $k/q^2$, which is typically many hundreds of micrometers or more. Therefore, in the integrand it is permissible to replace each $\rho(q,p,z,t)$ by its mean value, \emph{i.e.} by its zero Fourier component divided by the sample thickness, \emph{i.e.} $\rho(q,p,z,t) \simeq \tilde{\rho}(q,p,0,t)/\Lambda$. This factor, which is now independent of $z$, may be then taken outside of the integral, yielding
\begin{align}\label{eq:iqq}
     & \tilde I(q,p,t) \simeq {\left| {{a_0}} \right|^2}2{r_0} \\ \nonumber
    & \left( \int_\Lambda  dz {e^{\frac{{ - {{(z' - z)}^2}}}
{2}\left( {\frac{{{q^2}{\xi_x ^2}}}
{{{{(z' - z)}^2} + {k^2}{\xi_x ^4}}} + \frac{{{p^2}{\xi_y ^2}}}
{{{{(z' - z)}^2} + {k^2}{\xi_y ^4}}}} \right)}}\tfrac{{\xi_x \xi_y }}
{{{{({{(z' - z)}^2} + {k^2}{\xi_x ^4})}^{1/4}}{{({{(z' - z)}^2} + {k^2}{\xi_y ^4})}^{1/4}}}}   \right.\\ \nonumber
& \left(  \rho '(q,p,0,t)    \sin \left[  \Phi \right]  +  \rho ''(q,p,0,t)   \left. \cos \left[ \Phi \right] \right) \right)
\end{align}
Changing variable $s\equiv z'-z$, Eq.~\ref{eq:iqq} becomes
\begin{align}\label{eq:iqq2}
   &\tilde{I}(q,p,t)\simeq|a_{0}|^{2}\tfrac{2r_{0}}{k}\left(-\tilde{\rho}'(q,p,0,t)\tfrac{1}{\Lambda}\right.\\ \nonumber
   &\int_{z'}^{z'-\Lambda}dse^{-\frac{s^{2}}{2}\left(\frac{q^{2}\xi_x^{2}}{k^{2}
   \xi_x^{4}+s^{2}}+\frac{p^{2}\xi_y^{2}}{k^{2}\xi_y^{4}+s^{2}}\right)} \tfrac{\xi_x\xi_y}
   {(s^{2}+k^{2}\xi_x^{4})^{1/4}(s^{2}+k^{2}\xi_y^{4})^{1/4}} \\ \nonumber
   &\sin\left[\tfrac{1}{2}\left(-\tfrac{kq^{2}s\xi_x^{4}}{s^{2}+k^{2}\xi_x^{4}}-\tfrac{kp^{2}s
   \xi_y^{4}}{s^{2}+k^{2}\xi_y^{4}}+\tan^{-1}\tfrac{s}{k\xi_x^{2}}+\tan^{-1}\tfrac{s}{k\xi_y^{2}}\right)\right] \\ \nonumber
   &-\tilde{\rho}''(q,p,0,t) \tfrac{1}{\Lambda} \int_{z'}^{z'-\Lambda}dse^{-\frac{s^{2}}{2}\left(\frac{q^{2}
   \xi_x^{2}}{k^{2}\xi_x^{4}+s^{2}}+\frac{p^{2}\xi_y^{2}}{k^{2}\xi_y^{4}+s^{2}}\right)}\tfrac{\xi_x\xi_y}{(s^{2}
   +k^{2}\xi_x^{4})^{1/4}s^{2}+k^{2}\xi_y^{4})^{1/4}}\\ \nonumber
   &\left.\cos\left[\tfrac{1}{2}\left(-\tfrac{kq^{2}s\xi_x^{4}}{s^{2}+k^{2}\xi_x^{4}}-\tfrac{kp^{2}s
   \xi_y^{4}}{s^{2}+k^{2}\xi_y^{4}}+\tan^{-1}\tfrac{s}{k\xi_x^{2}}+\tan^{-1}\tfrac{s}{k\xi_y^{2}}\right)\right]\right)
\end{align}
The sample thickness $\Lambda$ is about the diameter of the capillary equal to 0.7 mm, which is much smaller than the sample-to-detector distance $s$ ranging from 53 mm to 203 mm. Thus, we assume that the integrand varies negligibly within the range of $z$ of the sample. As a result, Eq.~\ref{eq:iqq2} can be further simplified as:
\begin{align}\label{eq:iqq3}
   \tilde{I}(q,p,t)&\simeq|a_{0}|^{2}\tfrac{2r_{0}}{k}\left(\tilde{\rho}'(q,p,0,t)\right.\\ \nonumber
   &e^{-\frac{s^{2}}{2}\left(\frac{q^{2}\xi_x^{2}}{k^{2}
   \xi_x^{4}+s^{2}}+\frac{p^{2}\xi_y^{2}}{k^{2}\xi_y^{4}+s^{2}}\right)} \tfrac{\xi_x\xi_y}
   {(s^{2}+k^{2}\xi_x^{4})^{1/4}(s^{2}+k^{2}\xi_y^{4})^{1/4}} \\ \nonumber
   &\sin\left[\tfrac{1}{2}\left(-\tfrac{kq^{2}s\xi_x^{4}}{s^{2}+k^{2}\xi_x^{4}}-\tfrac{kp^{2}s
   \xi_y^{4}}{s^{2}+k^{2}\xi_y^{4}}+\tan^{-1}\tfrac{s}{k\xi_x^{2}}+\tan^{-1}\tfrac{s}{k\xi_y^{2}}\right)\right] \\ \nonumber
   &+\tilde{\rho}''(q,p,0,t) e^{-\frac{s^{2}}{2}\left(\frac{q^{2}
   \xi_x^{2}}{k^{2}\xi_x^{4}+s^{2}}+\frac{p^{2}\xi_y^{2}}{k^{2}\xi_y^{4}+s^{2}}\right)}\tfrac{\xi_x\xi_y}{(s^{2}
   +k^{2}\xi_x^{4})^{1/4}s^{2}+k^{2}\xi_y^{4})^{1/4}}\\ \nonumber
   &\left.\cos\left[\tfrac{1}{2}\left(-\tfrac{kq^{2}s\xi_x^{4}}{s^{2}+k^{2}\xi_x^{4}}-\tfrac{kp^{2}s
   \xi_y^{4}}{s^{2}+k^{2}\xi_y^{4}}+\tan^{-1}\tfrac{s}{k\xi_x^{2}}+\tan^{-1}\tfrac{s}{k\xi_y^{2}}\right)\right]\right)
\end{align}
\begin{figure}
\caption{The FWHM $w_q$ of the Gaussian distribution for the effect partial coherence of the source beam, plotted versus (a) sample-to-detector distance for different coherence length and (b) coherence length for different sample to detector distances. In (a), dashed lines denote the boundaries separating far-field and near-field for 8 $\rm{\mu m}$ and 163 $\rm{\mu m}$ from left to right. In (b), the intersections of dashed lines and color lines denote the differences of the $\sigma_q$ at different $s$ for $\xi = 8~\rm{\mu m}$ and $163~\rm{\mu m}$.}.\label{fig:coherence}
\includegraphics[width=0.8\textwidth]{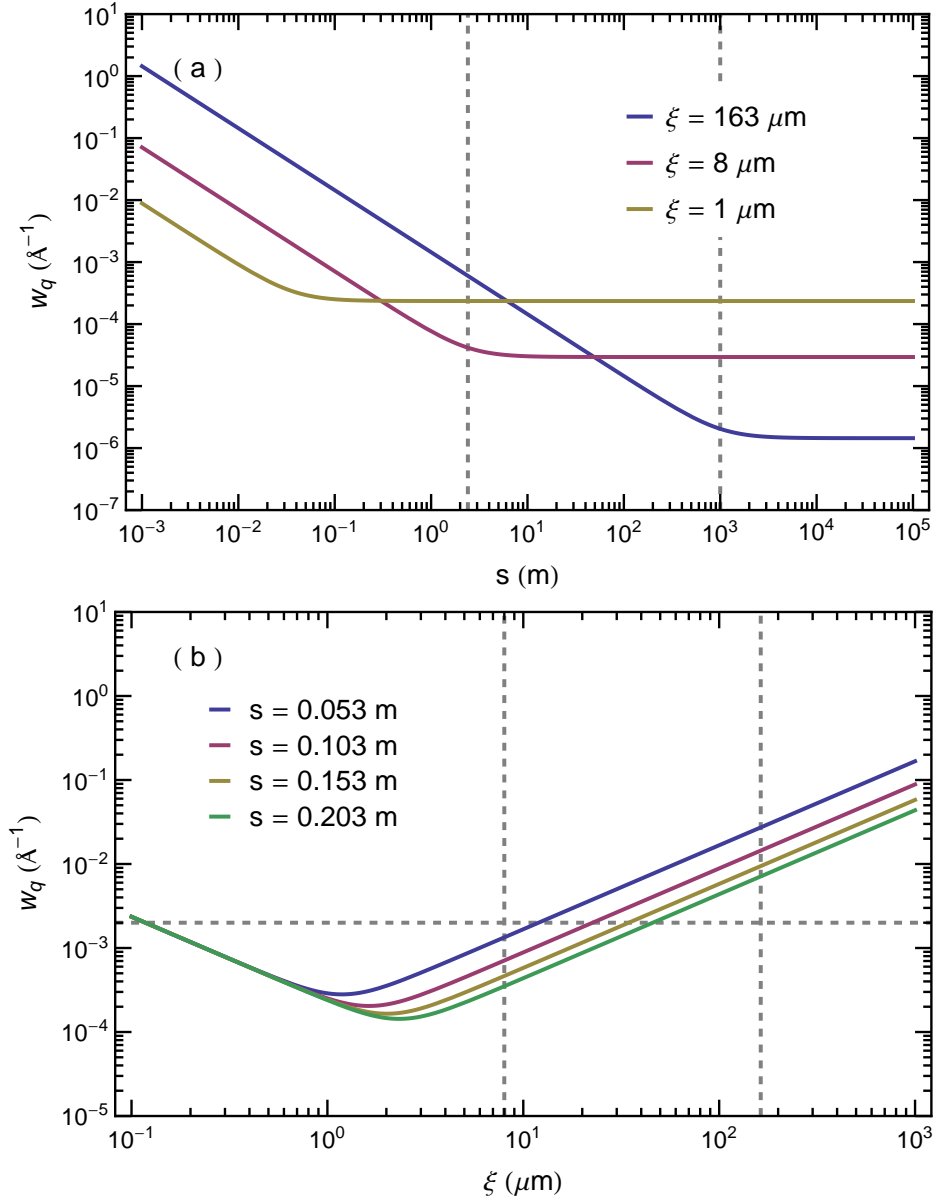}
\end{figure}

Here the term $e^{-\frac{s^{2}}{2}\left(\frac{q^{2}\xi_x^{2}}{k^{2}\xi_x^{4}+s^{2}}+\frac{p^{2}\xi_y^{2}}{k^{2}\xi_y^{4}+s^{2}}\right)}$ describes the effect of the partial coherence of the source beam on the speckle intensity. It is a product of two Gaussians with variances $\sigma_{q_{x;y}} = \sqrt{\frac{k^2\xi^4+s^2}{s^2\xi_{x;y}^2}}$. The full width at half maximum (FWHM) of the Gaussian distribution is $w_{q_{x;y}} = 2\sqrt{2\ln 2}\sigma_{q_{x;y}}$. Fig.~\ref{fig:coherence} plots the FWHM of coherence of the source beam versus (a) sample-to-detector distance $s$ and (b) coherence length $\xi$. From Fig.~\ref{fig:coherence}(a), we observe that $w_q$ decreases as $s$ increases until it reaches a constant value at a certain distance:
\begin{equation}
s_0 = k\xi^2,
\end{equation}
which is the usual near-field condition, called Fresnel condition. However, for XNFS that requires that the way scattered radiation falls onto the sensor duplicates the actual angular distribution of the scattered intensity, a much stronger condition should be satisfied~\cite{PhysRevLett.85.1416}:
\begin{equation}
s'_0 = k\xi a,
\end{equation}
where $a$ is the size of the scattering particle. With this condition, the source beam fills the whole field of view and the speckle size is related to the actual size of the probing material. In Fig.~\ref{fig:coherence}(b), $w_q$ is plotted versus coherence length for different values of $s$. For the estimated coherence lengths at 8ID indicated via the dashed lines in (b), in principle, one expects to observe an $s$-dependent change in the 2D speckle intensity. However, in reality, due to the limited spatial resolution, which in turn limits the $q$ range, and more critical, the sensor response~\cite{PhysRevLett.103.194805}, we have not been able to observe this $s$-dependent variation, as we discuss in more detail below. When $s^2 \ll k^2\xi_x^4$ and $s^2 \ll k^2\xi_y^4$ as we expect at 8-ID, then, Eq.~\ref{eq:iqq3}~may be simplified even further:
\begin{align}\label{eq:I(q)sim}
&\tilde{I}(q,p,t)\simeq|a_{0}|^{2}\frac{2r_{0}}{k}e^{-\frac{s^{2}}{2}\left(\frac
{q^{2}}{k^{2}\xi_x^{2}}+\frac{p^{2}}{k^{2}\xi_y^{2}}\right)}\\ \nonumber
&\left(\tilde{\rho}'(q,p,0,t) \right.\sin\left[\tfrac{1}{2}\left(-\tfrac{(q^{2}+p^{2})s}{k}+\tfrac{s}{k}\left(
\tfrac{1}{\xi_x^{2}}+\tfrac{1}{\xi_y^{2}}\right)\right)\right ]\\ \nonumber
&+\tilde{\rho}''(q,p,0,t) \left.\cos\left[\tfrac{1}{2}\left(-\tfrac{(q^{2}+p^{2})s}{k}+\tfrac{s}{k}
\left(\tfrac{1}{\xi_x^{2}}+\tfrac{1}{\xi_y^{2}}\right)\right)\right]\right),
\end{align}
  where the term $e^{-\frac{s^{2}}{2}\left(\frac{q^{2}}{k^{2}\xi_x^{2}}+\frac{p^{2}}{k^{2}\xi_y^{2}}\right)}$ is introduced as the spatial coherence transfer function in Ref.~\cite{cerbino:natphys}.

Introducing $\varphi =  \tan^{-1}(\tilde{\rho}'/\tilde{\rho}'')$, which is always small, except at x-ray energies near an absorption edge, and $\tilde \rho = \sqrt{(\tilde{\rho}')^2 + (\tilde{\rho}'')^2}$, we may re-write Eq.~\ref{eq:I(q)sim} as
\begin{align}\label{eq:I(Q)ofNFS}
{\tilde I}(q,p,t) &\simeq  I_0 \frac{2 r_0}{k} \tilde{\rho}(q,p,0,t) e^{\frac{-s^2}{2} \left(  \frac{q^2 }{k^2 \xi_x ^2} +\frac{p^2}{k^2 \xi_y ^2}  \right ) } \\ \nonumber
&{\sin} \left [ \tfrac{1}{2} \left ( - \tfrac{ q^2 s}{k} -\tfrac{ p^2 s}{k}
+ \tfrac{s}{k \xi_x ^2} + \tfrac{s}{k \xi_y ^2}  + \varphi \right ) \right ]   \\ \nonumber
&= I_0 \frac{2 r_0}{k} \tilde{\rho}(q,p,0,t) T(p,q)
\end{align}
where $p$ and $q$ are the wavevectors obtained in the $x-$ and $y-$directions, respectively, by numerically Fourier transforming the CCD image, $I_0 = |a_0|^2$, $\Lambda$ is the sample thickness, $\tilde{\rho}$ is the electron density in Fourier space, and where the latter equality defines the transfer function $T(q,p)$. It is worth emphasizing that the transfer function $T(\vec{Q})$ with $\vec{Q} = (q,p)$ is written as
\begin{align}\label{eq:T(Q)sim}
    T(\vec{Q})= e^{\frac{-s^2}{2} \left ( \right . \frac{q^2 }{k^2 \xi_x ^2} +\frac{p^2}{k^2 \xi_y ^2} \left. \right ) } {\sin} \left [ \tfrac{1}{2} \left ( - \tfrac{ Q^2 s}{k} + \tfrac{s}{k \xi_x ^2} + \tfrac{s}{k \xi_y ^2}  + \varphi \right ) \right ]
\end{align}

Eq. \ref{eq:I(Q)ofNFS} immediately allows us to calculate the static structure factor $S(\vec Q)$ in terms of measured quantities. Specifically,
\begin{equation}\label{eq:S(q,p)}
    S(\vec Q ,0) = \left\langle{\left|\rho(q,p,0,t)\right|^2}\right\rangle = \frac{\left\langle {\left| {I(\vec Q,t)} \right|^2 } \right\rangle}{ (4 r_0^2 I_0^2 |T(\vec Q)|^2 /k^2)}
\end{equation}
Similarly, it is straightforward to show that, in the context of XNFS,
the normalized IFS is
\begin{align}\label{eq:g1(q,p)}
    {g_1}(\vec{Q},\tau ) &= \frac{{\left\langle {\rho (q,p,0,t){\rho ^ * }(q,p,0,t + \tau )} \right\rangle }}
{{\left\langle {{{\left| {\rho (q,p,0,t)} \right|}^2}} \right\rangle }}  \\ \nonumber
&= \frac{{\left\langle {I(\vec{Q},t){I^ * }(\vec{Q},t + \tau )} \right\rangle }}{{\left\langle {{{\left| {I(\vec{Q},t)} \right|}^2}} \right\rangle }}
\end{align}

According to Eq.~\ref{eq:g1(q,p)}, $g_1$ is independent of the transfer function $T(\vec Q)$, and in turn does not depend on sample-to-detector distance, $s$.

In summary, the intensity measured in the near field speckle experiments is proportional to the density of the sample rather than the modulus squared of the density as in conventional far field speckle experiments like XPCS and SAXS. Thus, the time autocorrelation of $I(Q)$ gives directly the intermediate scattering function $g_1$ (Eq.~\ref{eq:g1(q,p)}). When the delay time $\tau$ is chosen to be zero, we obtained a quantity proportional to the static structure factor $S(Q)$ (Eq.~\ref{eq:S(q,p)}) times the NFS transfer function $|T(Q)|^2$.

\section{Detector design}

High spatial resolution and high detection efficiency are key goals for imaging the x-ray speckles in XNFS experiments. In XNFS experiments, in order to resolve micron-sized particles, it is necessary to employ detectors capable of resolving micrometers. A typical x-ray imaging detector consists of a crystal x-ray scintillator, a microscope objective and a fast, high-resolution, large-dynamic-range charge-coupled device (CCD)-based camera. The scintillator converts x-rays into visible light; the objective collects the visible light and magnifies the visible-light image; and finally the CCD camera records the image. Our aim was to design a detector with the best combination of scintillator and objective to achieve the optimal combination of spatial resolution and detection efficiency for XNFS experiments.

A key characteristic of an objective is its numerical aperture (NA). The numerical aperture (NA) of an objective defines the largest angle of light acceptance as well as the light collecting power. The detection efficiency scales as (NA)$^2$. Thus, a large NA objective is necessary for high detection efficiency. It is common to use immersion oil of high refractive index (n = 1.515) between the front lens of the objective and the scintillator to achieve high numerical aperture. We employed an Nikon Plan Fluor 40x oil immersion microscope objective with a numerical aperture of NA = 1.3, a working distance of WD = 0.2 mm and a field of view of diameter 0.67 mm. This objective uses an infinity focused optical system with a reference focal length of 200 mm. In our case, with the implementation of a tube lens with adjustable focal length from 25 mm to 150 mm, the 40x objective gives a real magnification of from 5 to 30 times.

\begin{figure}
    \includegraphics[width=0.8\textwidth,keepaspectratio=true]{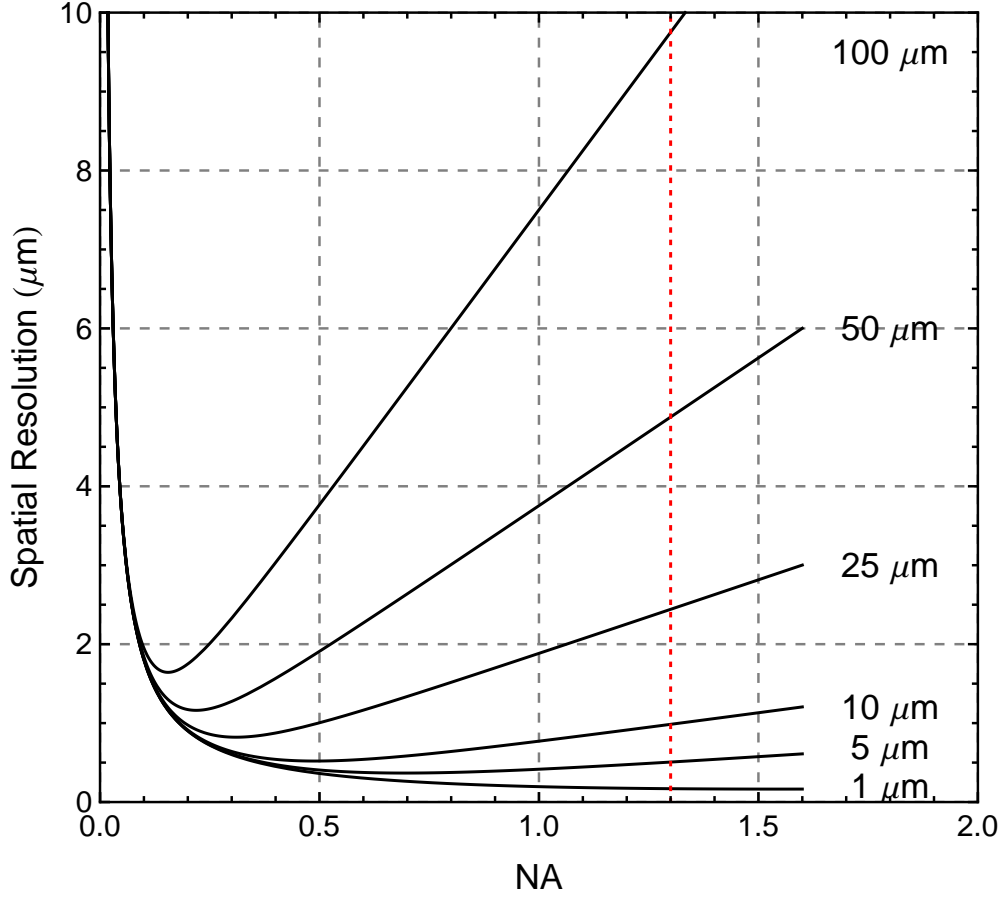}
    \caption{Spatial resolution ($\rm{\mu m}$) versus numerical aperture NA of the objective for different x-ray absorption lengths of the scintillator adapted from Ref.~\cite{Martin:gf0006}.}
    \label{fig:resolution}
\end{figure}

Generally, taking into account both the effects of diffraction and depth of focus, the spatial resolution [$R$] as a function of NA is given by~\cite{Martin:gf0006} :
\begin{equation}\label{eq:resolution}
    R = [(p/ \mbox{NA} )^2  + (q\Delta z \mbox{NA} )^2 ]^{1/2}
\end{equation}
where, $p=0.18$ and $q=0.075$ are constants obtained by numerical simulations~\cite{Martin:gf0006} and $\Delta z$ is the x-ray absorption length of the scintillator. Based on Eq. ~\ref{eq:resolution}, one can plot $R$ versus NA for different $\Delta z$, as shown in Fig.~\ref{fig:resolution}. It is clear from Fig.~\ref{fig:resolution} that to achieve a spatial resolution of a micrometer or less with high detection efficiency (NA $\geq1.0$) , we have to choose a scintillator with x-ray absorption length of 10 $\rm{\mu m}$ or less.

Besides the x-ray absorption length, several additional characteristics of the crystal scintillator are critical for x-ray imaging, including: high x-ray stopping power; high light yield; the emission wavelength being compatible with CCD readout (400 nm - 700 nm); a similar refractive index to the immersion oil (n=1.515), and a small thickness to minimize spherical aberration. At a minimum, the scintillator thickness should be smaller than the working distance of the objective (0.2 mm), so that the objective can focus to the upstream side of the scintillator.

\begin{table}
\caption{Characteristics of the scintillators}
\begin{tabular}{p{2cm}p{0.8cm}p{1cm}p{0.8cm}p{1cm}p{1cm}}
\toprule \toprule
scintillator & x-ray absorption length ($\rm{\mu m}$) & light yield per keV & refractive index & density ($\rm{g\cdot cm^{-3}}$) & Wavelength of max. emission (nm)\\
\midrule
LYSO (Lu$_{1.8}$Y$_{0.2}$SiO$_{5}$:Ce) & 9.5 & 32 & 1.81 & 7.1 & 420\\
YAG:Ce (Y$_3$Al$_5$O$_{12}$:Ce) & 25 & 8 & 1.82 & 4.55 &550\\
CdWO4   & 6.5 & 12 to 15 & 2.2 & 7.9 & 475\\
\bottomrule
\end{tabular}\label{tb:scint}
\end{table}

As shown in Table~\ref{tb:scint}, one potential candidate is YAG:Ce (Y$_3$Al$_5$O$_{12}$:Ce) which has been widely used in x-ray imaging detectors. However, its x-ray absorption length at 7.44 keV is about 25 $\rm{\mu m}$, which is not suitable for submicrometer resolution detector with a NA = 1.3 objective (Fig.~\ref{fig:resolution}). Another candidate is CdWO$_4$ whose x-ray absorption length at 7.44 keV is only 6.5 $\rm{\mu m}$. However, It is very difficult to obtain thin crystals of CdWO$_4$. It easily breaks before being thinned to the desired thickness ($<0.2$ mm) due to its (010) cleavage plane. In addition, the CdWO$4$ refractive index (n=2.2) is much different from that of immersion oil (n=1.515), which will induce a relatively large spherical abberation. In our setup, we use LYSO ($\rm{Lu_{1.8}Y_{0.2}SiO_2}$) which appears to be the most appropriate candidate overall. Its x-ray absorption length at 7.44 keV is 9.5 $\rm{\mu m}$, slightly larger than the x-ray absorption length of CdWO$_4$ but still good enough to produce high spatial resolution. Its light yield is 32 photons per keV, much higher than both CdWO$_4$ and YAG:Ce. Its refractive index (n=1.81) is not too far from that of immersion oil inducing less spherical abberation than CdWO$_4$. In addition, we found a manufacturer providing two-sided polished LYSOs with a thickness of 0.15 mm, just appropriate for our objective, although, as we show below, even thinner would lead to reduced spherical aberrations.

Here, following Koch \cite{koch98}, we briefly discuss spherical aberrations of our system. Generally, ``spherical aberration'' defines the aberration that is introduced when the light from near the center of the lens has a different focal length from light from the margins of the lens. We begin with calculating the spherical aberration induced by a general material with refractive index of $n_1$ and a thickness of $t$. According to snell's law, an incident ray with an angle of $\theta_1$ against the optical axis will refract at the interface of the material and another media with refractive index $n_2$ with a new angle $\theta_2$:
\begin{equation}\label{eq:snell}
    n_1 \sin \theta_1 = n_2 \sin \theta_2
\end{equation}
Then, the distance from the virtual source of ray and the interface is given by
\begin{equation}
d = t\frac{\tan \theta_1}{\tan\theta_2} =t\frac{n_{2}}{n_{1}}\frac{\cos\theta_{2}}{\cos\theta_{1}}
\end{equation}
So the shift of the focus is given by the difference between $d$ for the paraxial ray and marginal ray:
\begin{align}
    \Delta z &= d(\theta_1,\theta_2 = 0) - d(\theta_1,\theta_2) = t \frac{n_{2}}{n_{1}} \left(1 - \frac{\cos \theta_2}{\cos \theta_1}\right) \\ \nonumber
    &=t\frac{n_{2}}{n_{1}}\left(1-\frac{\sqrt{1-\sin\theta_{2}^{2}}}{\sqrt{1-\frac{n_{2}^{2}}{n_{1}^{2}}\sin\theta_{2}^{2}}}\right) \\ \nonumber
    &=t\frac{n_{2}}{n_{1}}\left(1-\frac{\sqrt{1-\sin\theta_{2}^{2}}}{\sqrt{1-\sin\theta_{2}^{2}+(1-\frac{n_{2}^{2}}{n_{1}^{2}})\sin\theta_{2}^{2}}}\right)\\\nonumber
    &=t\frac{n_{2}}{n_{1}}\left(1-\frac{1}{\sqrt{1+(1-\frac{n_{2}^{2}}{n_{1}^{2}})\frac{\sin\theta_{2}^{2}}{1-\sin\theta_{2}^{2}}}}\right) \\ \nonumber
    &\approx t\frac{n_{2}}{n_{1}}\left(1-1+\frac{1}{2}(1-\frac{n_{2}^{2}}{n_{1}^{2}})\frac{\sin\theta_{2}^{2}}{1-\sin\theta_{2}^{2}}\right)\\ \nonumber
    & =\frac{1}{2}t\frac{n_{2}n_{1}^{2}-n_{2}^{3}}{n_{1}^{3}}\sin\theta_{2}^{2}+O(\sin\theta_{2}^{4}) \\ \nonumber
    & \approx \frac{1}{2}t\frac{n_{2}n_{1}^{2}-n_{2}^{3}}{n_{1}^{3}}\frac{a^2\rho'^2}{r^2}
\end{align}
where we used $\tan \theta_1 / \tan \theta_2 = d / t$, $t$ is the thickness of the scintillator, $\sin \theta_2 \approx \tan \theta_2 = \frac{a\rho'}{r}$, $a$ is the aperture of objective and $\rho'$ is the normalized radius of the ray on the entrance pupil. Following the addition theorem for the primary aberrations~\cite{born:1999} (p246): \emph{Each primary aberration coefficient of a centered system is the sum of the corresponding aberration coefficients associated with the individual surfaces of the system}, we can replace $\rho'$ by the radius of the ray on the exit pupil $\rho$, and obtain the wave aberration due to shift of focus: \cite{born:1999} (p550):
\begin{equation}\label{eq:psish}
    \psi(q) = \frac{1}{2}\left(\frac{a}{r}\right)^2\Delta z \rho^2 = \frac{1}{2}\left(\frac{a}{r}\right)^2 \rho^2 \left(\frac{1}{2}t\frac{n_{2}n_{1}^{2}-n_{2}^{3}}{n_{1}^{3}}
 \frac{a^2\rho^2}{r^2}\right)
\end{equation}
And the spherical aberration is represented by \cite{born:1999} (p530):
\begin{equation}
    \psi(q)_{sp} = A'_{040}\rho^4
\end{equation}
Hence, from Eq~\ref{eq:psish}. we find that $A'_{040}$ is of the form:
\begin{equation}\label{eq:A040}
    A'_{040}=\frac{1}{2}\mathrm{NA}^4t\frac{n_{2}n_{1}^{2}-n_{2}^{3}}{2n_{1}^{3}}
\end{equation}
where we use $\mathrm{NA} = a/r$.
Born and Wolf \cite{born:1999} also derive the tolerance condition of the spherical aberration:
\begin{equation}
    A'_{040} < 0.94 \lambda
\end{equation}
Hence, the maximum tolerated thickness of the material satisfies
\begin{equation}\label{eq:tolerance}
    t < 0.94 \lambda / \left(\frac{1}{2}\mathrm{NA} ^4 \frac{n_{2}n_{1}^{2}-n_{2}^{3}}{2n_{1}^{3}} \right)
\end{equation}
For $n_1 = 1.515$, $n_2=1$ and $\lambda = 420$ nm, the tolerated thickness is 1.5 $\rm{\mu m}$, which is too hard to reach. So the correction of spherical aberration is required. The objective we used is an oil-immersion objective with a working distance of 0.2 mm that is corrected for glass coverslips of a thickness of 0.17 mm. The spherical aberration induced by using a different refractive index $n'$ can be compensated by choosing a different thickness $t'$, which is given by\cite{koch98}
\begin{equation}\label{eq:thinkness}
    t' = \frac{n_1^2-1}{n_1^3} \frac{n'^3}{n'^2-1}t
\end{equation}
When working with a crystal scintillator of a refractive index of 1.81 and a thickness of 0.15 mm, the equivalent thickness of the glass and oil is 0.155 mm. So what we need to do is remove the coverslip and increase the oil thickness to 0.215 mm, then the resultant spherical aberration is within the tolerance.

The lateral displacement of the image for the spherical aberration is given by ~\cite{born:1999} (p531):
\begin{equation}
    \Delta Y = 4 \left(\frac{R}{a}\right)\left(\frac{Y}{a}\right)^3A'_{040}
\end{equation}
Then the resolution due to the spherical aberration is proportional to the lateral displacement of the image
\begin{equation}\label{eq:dx}
    R \propto \Delta y_{max} = \frac{\sin \alpha }{\mathrm{NA}} \Delta Y_{max} = \frac{4}{\mathrm{NA}}A'_{040} \approx t \mathrm{NA}^3
\end{equation}
where $\Delta Y_{max}$ is obtained when $Y = a$, and we used $\sin \alpha = a/R$ and the sine condition $ y\cdot n\sin\theta = Y \sin \alpha$. So the contribution of the spherical aberration is not trivial for high numerical aperture objective. To lessen this effect, one can either choose small NA objective, or choose a scintillator with thinner thickness.

\section{Experimental setup}

\begin{figure}
    \includegraphics[width=0.9\textwidth,keepaspectratio=true]{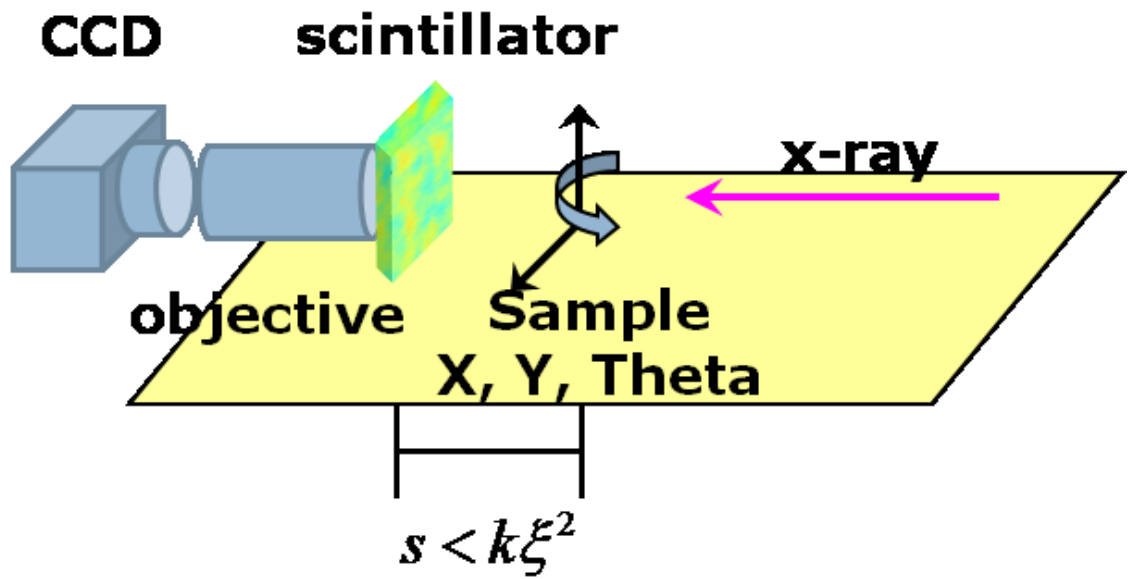}
    \caption{A sketch of the optical setup for XNFS experiment. The key components from right to left are: the beam source, the sample stage, the scintillator, the microscope objective and the CCD camera.}
    \label{fig:opticalsetup}
\end{figure}

X-ray near-field speckle (XNFS) measurements were carried out at beamline 8-ID-I of the Advanced Photon Source (APS) at Argonne National Laboratory, using x-rays of energy 7.44 keV and half the source beam size available at 8-ID-I hutch - about 0.5 $\times$ 0.5 $\rm{mm^2}$. Fig.~\ref{fig:opticalsetup} shows a sketch of the optical setup located in the 8-ID-I hutch. From right to left, we have the beam source, the sample stage, the scintillator, the microscope objective and the CCD camera. In XNFS experiments, no spatial and spectral filtering of the direct beam are required~\cite{cerbino:natphys}. We inherit the exiting setup for XPCS which gives an energy resolution of $\Delta E / E \approx 3 \times 10^{-4}$ and remove all the slits letting the full beam impinge onto the sample and record the interference pattern of the transmitted and scattered beams by means of our detector placed at distances from the sample ranging from $z = 53$ mm to $203$ mm.

As described before, we use two-sided polished LYSO ($\rm{Lu_{1.8}Y_{0.2}SiO_2}$) with a thickness of 0.15 mm to convert x-rays into visible light and a Nikon Plan Fluor 40x oil immersion microscope objective with a numerical aperture of NA = 1.3 to magnify the image. Both of them are mounted on a piezo electric stage which has a mechanical manual adjustable coarse travel range of 4 mm and a piezo electric travel range of 20 $\rm{\mu m}$ with a resolution of 20 nm. As a result, by carefully tuning the distance between the objective and the scintillator, we are able to focus the image which is in turn magnified 5 to 30 times by a tube lens with adjustable focal length and then recorded by a ``CoolSNAP'' CCD camera. The camera features a 1392 $\times$ 1040 pixels with size 6.45 $\rm{\mu m}$ $\times$ 6.45 $\rm{\mu m}$ and a maximum frame rate of 56 Hz. All the images so-obtained are subsequently cropped to 1024 $\times$ 1024 pixels for convenience of 2-D Fourier transformations in the data analysis.

There are several contributions to the detection resolution:
(1) The resolution of the scintillator -- for x-rays incident onto the scintillator at a single point. According to Refs. \cite{koch98,Martin:gf0006}, this is typically 0.1~$\rm{\mu} m$ for 7-8~keV x-rays, smaller than optical limits on the resolution. (2) The resolution determined by the diffraction limit and the defect of focus of the objective and the scintillator via Eq.~\ref{eq:resolution}. With an objective with NA = 1.3 and LYSO with an x-ray absorption length of 9.5 $\rm{\mu m}$, we obtain a spatial resolution of $0.98~\rm{\mu m}$. (3) The reduction in resolution caused by the spherical aberration due to the use of s scintillator of mismatched refractive index, which is proportional to the thickness of the scintillator and the cube of NA (Eq.~\ref{eq:dx}). Since our objective is corrected for spherical aberration and the additional spherical aberration induced by replacing the glass coverslip by the scintillator can be compensated for by adjusting the oil thickness , this factor is not critical to the spatial resolution. (4) The (demagnified) size of the CCD pixel, if too large, could limit the resolution. For a magnification of 30, the CoolSNAP can resolve a length scale as small as $6.45  /30 = 0.215~\rm{\mu m}$ which does not limit the resolution. As a result, the spatial resolution of detector should be largely determined by factor (2), which is about 1 $\rm{\mu m}$.

Hence, we are able to estimate the maximum $q$ range. With $r_{min} = 0.98~\rm{\mu m}$, in reciprocal space, $q_{max} = \frac{\pi}{R} \sim 4 \times 10^{-4}$ \AA $^{-1}$. On the other hand, the lower limit of the wavevector $q_{min}$ should be determined by the largest accessible length scale. In principle, the largest length scale is the size of the measured scattering image equal to $1024 \times 0.215~\rm{\mu m} = 0.22$ mm, so $q_{min} = \frac{2 \pi}{0.22 \rm{mm}} \sim 3 \times 10^{-6}$ \AA $^{-1}$. However, practically our data shows identical low $q$ profile that is independent of which sample is being studied and of the sample-to-detector distance. This is likely due to the beam structure on long length scales. So, the realistic useful $q_{min}$ is of the order of $10^{-5}$ \AA$^{-1}$. Nevertheless, the $q$-range achieved is at least a decade below the range accessible to the conventional XPCS experiments.

\section{Silica 0.45 ${\rm{\mu m}}$ suspension}

The first sample, we measured at 8-ID-I, as an initial test of our XNFS setup is a colloidal suspension of silica particles of 0.45 $\rm{\mu m}$ diameter and a volume fraction ($\phi$) around 0.05. The static structure factor peak of this sample is expected to be located around $10^{-3}$ \AA$^{-1}$, which means that the sample has a uniform scattering profile in the $q$ range accessible in our XNFS setup (\emph{i.e.} $S(q) \simeq const$). In other words, the intensity profile $I(q)$ we measured from this silica suspension should simply result from the transfer function $T(q)$ (Eq.~\ref{eq:I(Q)ofNFS}). Thus, it should be an ideal sample to examine the sample-to-detector distance ($s$) dependence of $T(q)$.

\begin{figure}
\centering
\includegraphics[width=1\textwidth]{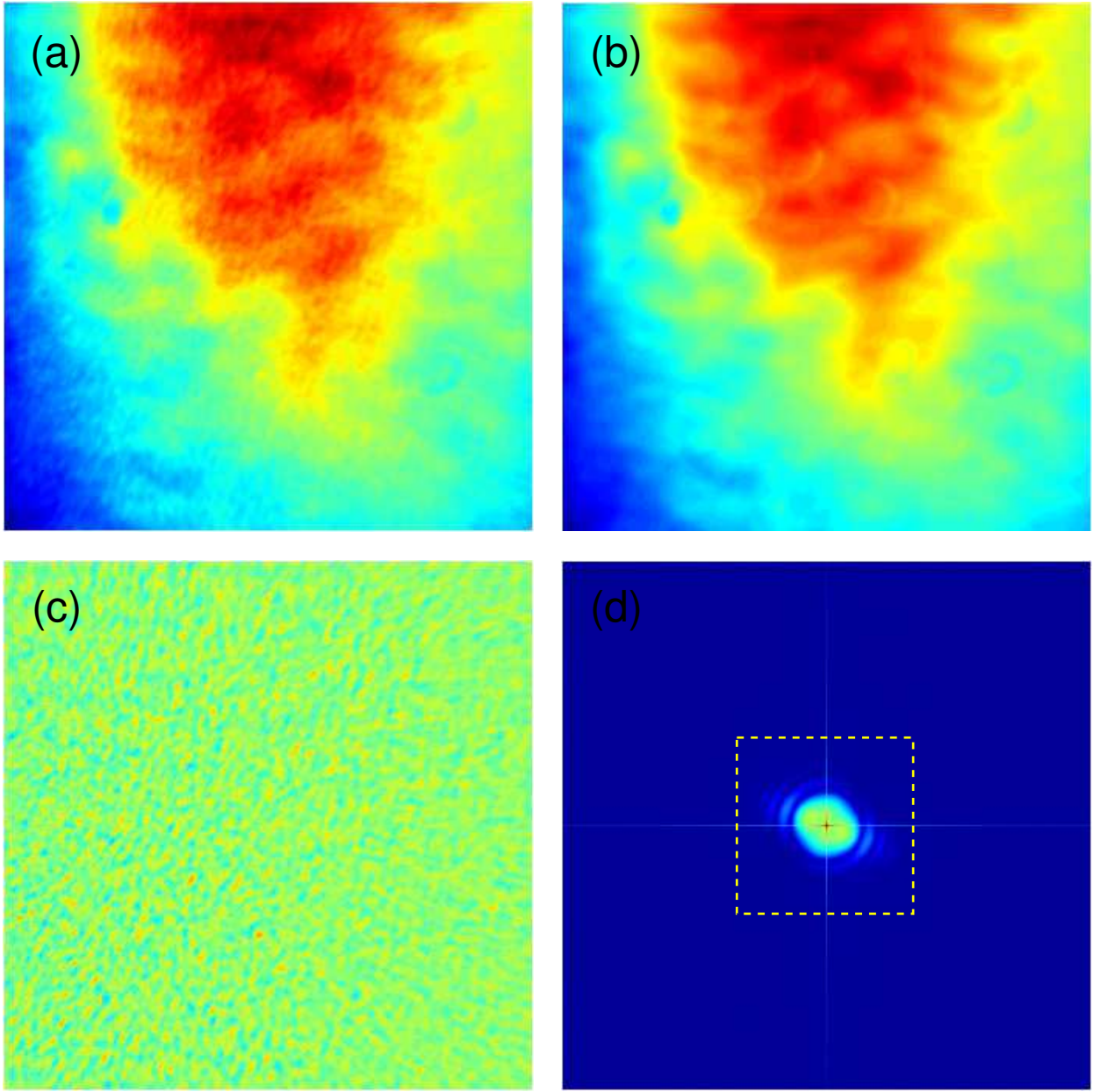}
\caption{(a) A raw single frame of scattering image of SiO$_2$ with diameter of 0.45 ${\rm{\mu m}}$ at sample-to-detector distance of 53 mm. (b) The image averaged over 1000 frames. (c) Normalized image obtained by dividing raw image with averaged image. Each image consists of 1024 $\times$ 1024 pixels. (d) Two-dimensional fourier transform of (c).}\label{fig:si0p45images}
\end{figure}

Illustrated in Fig.~\ref{fig:si0p45images}(a) is the typical raw image of the silica suspension measured at $z = 53$ mm. The image contains 1024 $\times$ 1024 pixels with a pixel size of $d_{pix} = 6.45~\rm{\mu m}$. The magnification of the detector is set to 30. Thus this image corresponds to a region of size 0.22 mm $\times$ 0.22 mm in the sample. The speckle pattern appears quite obscure and weak due to the large-scale, static background. After averaging over 1000 frames, the speckle pattern is washed away leaving the static beam profile fluctuation unchanged, as shown in Fig.~\ref{fig:si0p45images}(b). In order to remove the large-scale static fluctuations, we perform a normalization of each raw image by dividing each one with an average image, averaged over 1000 frames, as shown in Fig~\ref{fig:si0p45images}(b). Fig.~\ref{fig:si0p45images}(c) presents one example of the resultant image, which reveals a clear, uniform speckle pattern ($I(x,y)$). Next, a two-dimensional Discrete Fourier transform is performed via
\begin{equation}
I(q,p)=\frac{1}{N}\overset{N-1}{\underset{x=0}{\sum}}\overset{N-1}{\underset{y=0}{\sum}}I(x,y)\exp(-i(qx+py)/N)
\end{equation}
which produces the $q$-space image, as shown in Fig.~\ref{fig:si0p45images}(d). Here, N=1024. As a result, the corresponding $q$ coordinates are given by:
\begin{equation}
    q = \frac{2\pi}{N \Delta r}\frac {x}{\Delta r}\;;\;p = \frac{2\pi}{N \Delta r}\frac{y}{\Delta r}
\end{equation}
where $\Delta r$ is the size of one pixel, equal to $d_{pix}/30=0.215~\rm{\mu m}$.

\begin{figure}
\centering
\includegraphics[width=1\textwidth]{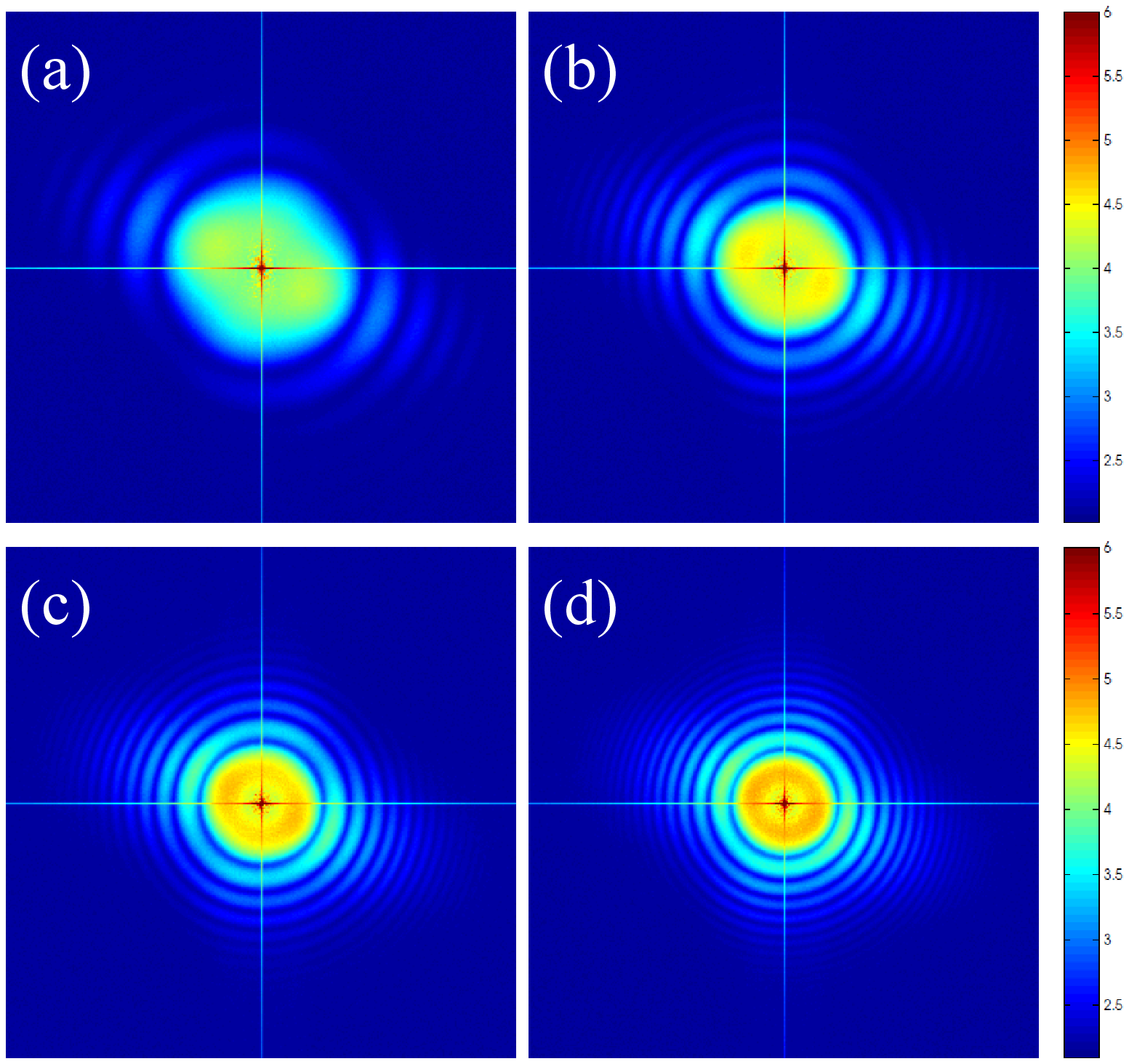}
\caption{Enlarged Fourier Transformed scattering image of SiO$_2$ suspension at different sample-to-detector distances (a) 53 mm (b) 103 mm (c) 153 mm (d) 203 mm.}\label{fig:si0p45FFT}
\end{figure}

We report in Fig.~\ref{fig:si0p45FFT} several examples of the magnified Fourier transformed image corresponding to the region inside the dashed lines in Fig.~\ref{fig:si0p45images}(d). Different panels are obtained at different sample-to-detector distances: (a) $53$ mm, (b) $103$ mm, (c) $153$ mm and (d) $203$ mm. In each image, prominant fringes can be seen. It is clear that the fringes become finer when the detector is moved away from the sample. This agrees with the theoretical prediction that the transfer function $|T(q)|^2$ is proportional to a sine term whose frequency depends on the sample to detector distance (Eq.~\ref{eq:T(Q)sim}). Note that the rings of the Fourier transformation are not azimuthally uniform. This effect may be related to asymmetry in the coherence of the source beam. However, the envelop of the asymmetry has no obvious $s$-dependence, in contrast to what may be expected on the basis of Eq.~\ref{eq:iqq3}. Thus, we do not understand this asymmetry in detail. In fact, examination of these data (Fig.~\ref{fig:si0p45FFT}), in the light of Eq.~\ref{eq:iqq3}, suggests that the predicted effect of a finite coherence length is not playing a role in determining these data, presumably because the width of the Gaussian in Eq.~\ref{eq:iqq3} is greater than our accessible $q$-range of $q_{max} \approx 2 \times 10^{-3}$ \AA$^{-1}$ even for the largest values of $s$ studied. In Ref.~\cite{PhysRevLett.103.194805}, the authors showed that by far the largest contribution to the $q$ decay of the speckle power spectrum is due to the sensor transfer function. It is highly possible that this asymmetry rings are due to the sensor response.

\begin{figure}
\centering
\includegraphics[width=1\textwidth]{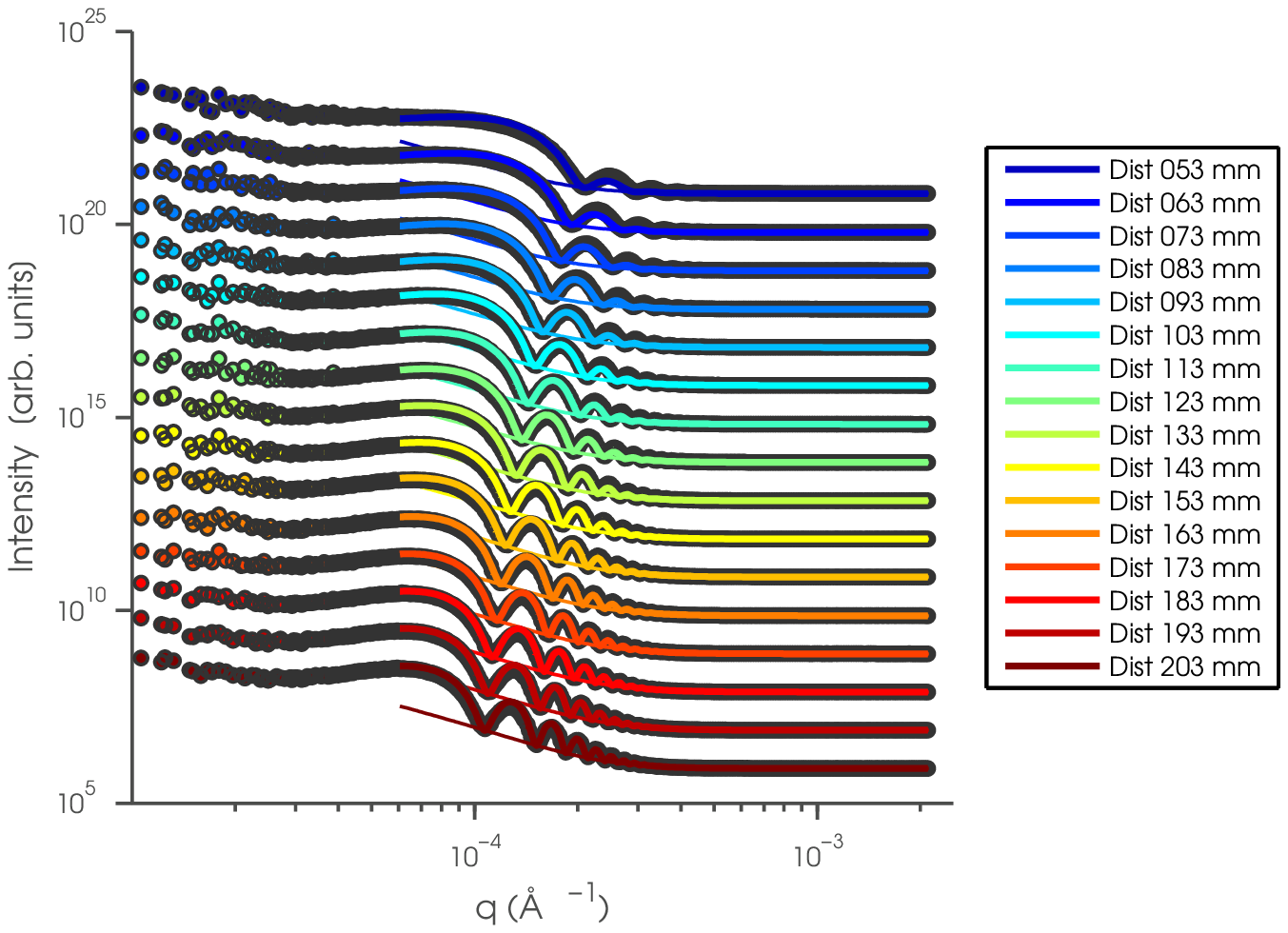}
\caption{Intensities of SiO$_2$ suspension at different sample-to-detector distances. Symbols are the data. Thick solid lines are the fittings based on Eq.~\ref{eq:I(q)forSiO2}. Thin solid lines are obtained by fitting the local minima with a stretched exponent, describing the contribution of multiple scattering. Data are displaced by a factor of 10 for clarity.}\label{fig:si0p45intfit}
\end{figure}

To quantify $|T(q)|^2$, we plot in Fig.~\ref{fig:si0p45intfit} the azimuthally-averaged intensity profile (symbols) versus wavevector ($q$) for different values of $s$ varying from $203$ mm to $53$ mm with a decrement of $10$ mm. The thick solid lines in Fig.~\ref{fig:si0p45intfit} are the fits of $I(q)$s to the relation:
\begin{align}\label{eq:I(q)forSiO2}
    I(Q) = A_1  \left| T(Q) \right|^2 + A_2 e^{-(v \cdot Q )^\beta}    + B
\end{align}
with  $Q = |\vec{Q}| = \sqrt{q^2 + p^2}$ and $T(Q)$ as a simplification of Eq.~\ref{eq:T(Q)sim}:
\begin{equation}\label{eq:T(Q)fit}
T(Q) = e^{-(w \cdot Q)} \cdot e^{\frac{-s^2}{2} \left(  \tfrac{q^2 }{k^2 \xi_x ^2} +\tfrac{p^2 }{k^2 \xi_y ^2}  \right) } \cdot {\sin} \left [  \left ( - \tfrac{ Q^2 s}{2k} + \tfrac{s}{k \xi ^2}   + \varphi \right )  \right ]
\end{equation}
which consists of a product of three terms. The first factor is additional to Eq.~\ref{eq:T(Q)sim} in order to take into account the limited spatial resolution of our optical system. As in Eq.~\ref{eq:T(Q)sim}, the second term derives from the partial coherence of the incident beam. How the coherence length enters is somewhat counterintuitive: It is proportional to the width of the Gaussian term in reciprocal space. However, this term is set to unity for fitting, since it plays no significant role in the accessible $q$ range, as we observed in the context of Fig.~\ref{fig:si0p45FFT}. The third term is a sine function that describes the fringes produced by the interference of the scattered beam and incident beam. In x-ray domain, this term is called phase contrast transfer function~\cite{cerbino:natphys}. Note we assume $\xi_x = \xi_y = \xi$ for the simplification of a phase factor in the sine function.

\begin{figure}
\centering
\includegraphics[width=0.9\textwidth]{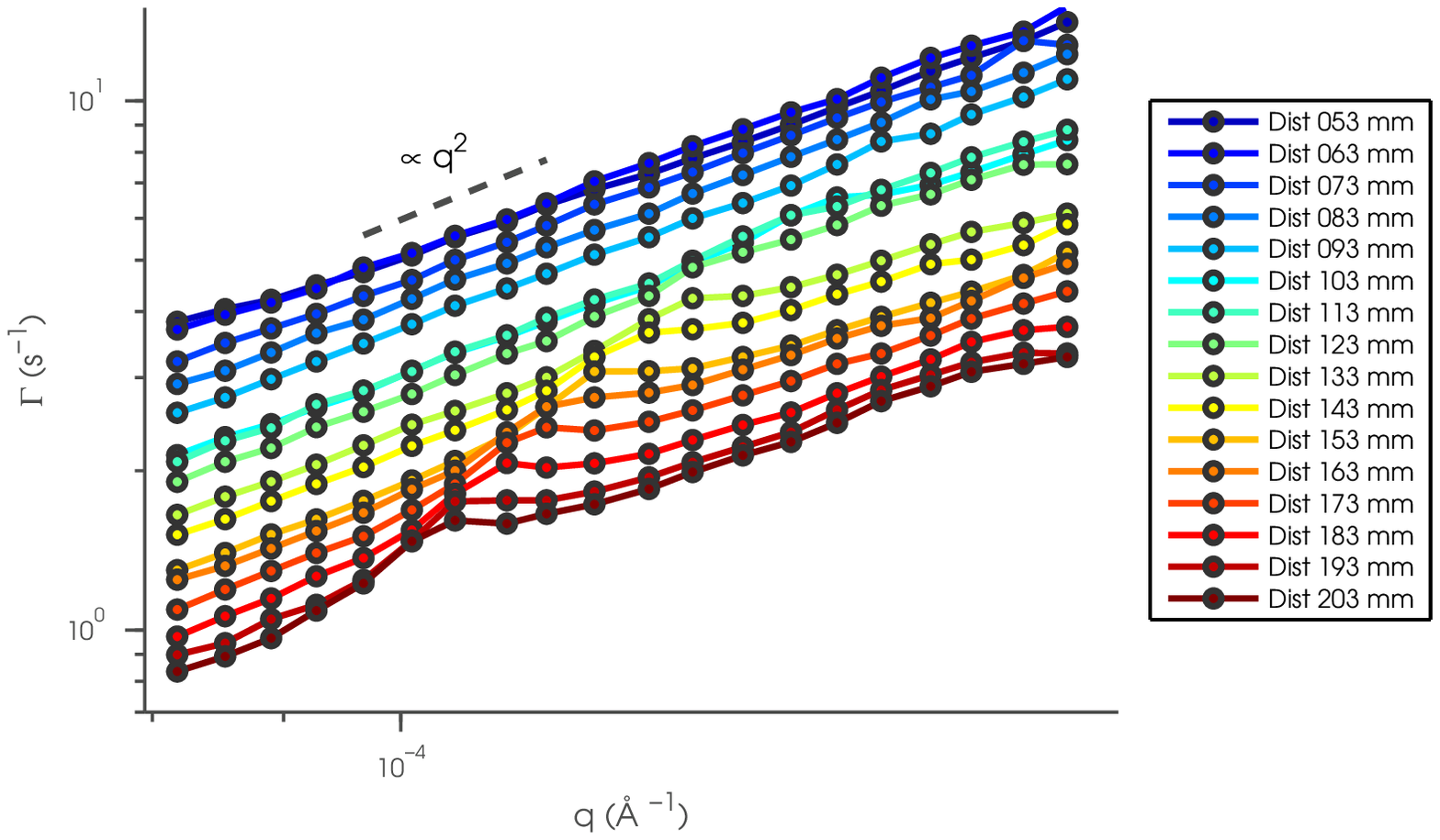}
\caption{The best-fit decay rate $\Gamma$ versus wavevector $q$ for silica of diameter 0.45 $\rm{\mu m}$ measured at different $s$. Points are the fitting results and the lines are guides to the eye. Data are displaced by a factor of 1.1 for clarity.}\label{fig:si0p45gamma}
\end{figure}

\begin{figure}
\includegraphics[width=1\textwidth]{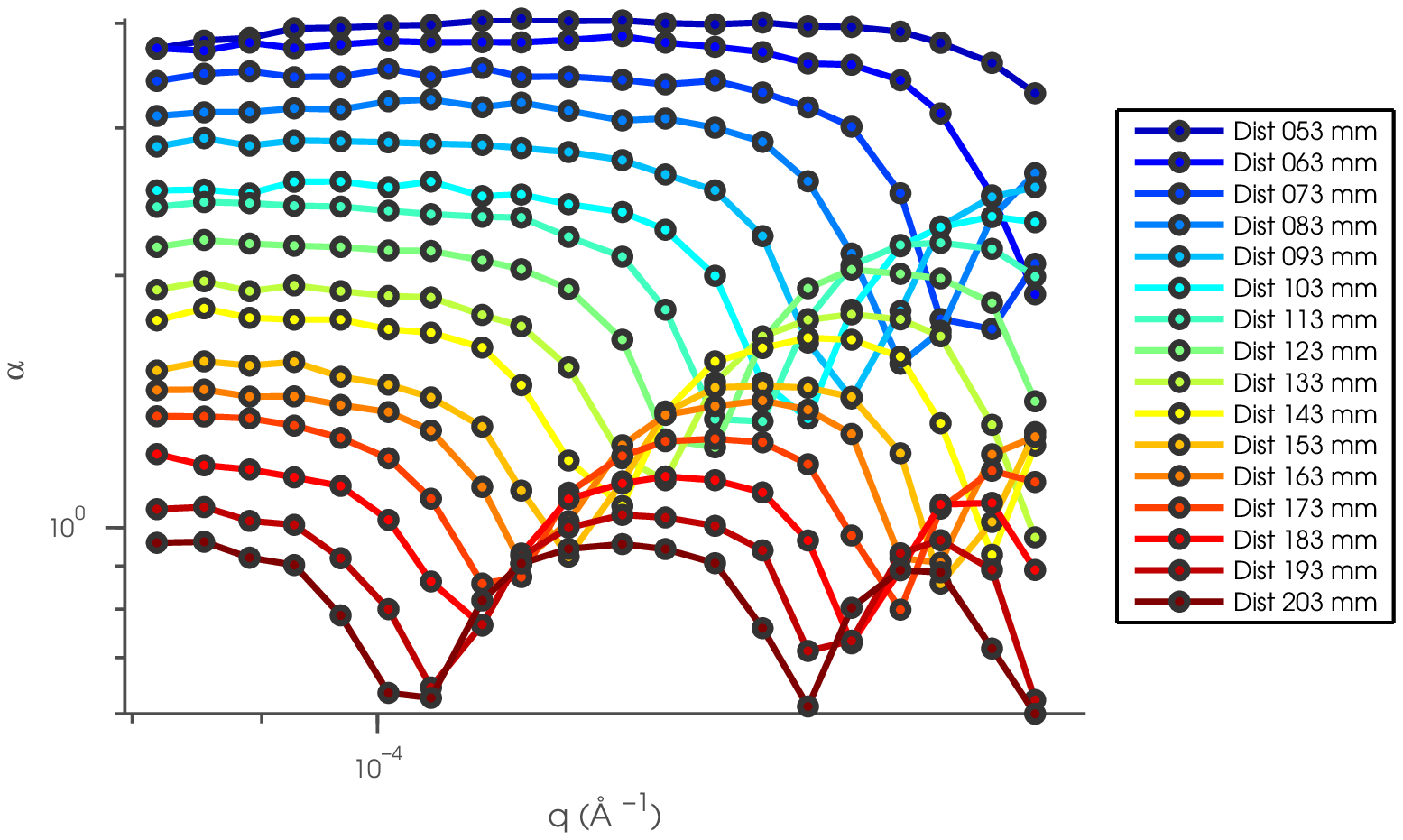}
\caption{The best-fit exponent $\alpha$ versus wavevector $q$ for silica of diameter 0.45 $\rm{\mu m}$ measured at different $s$. Points are the fitting results and the lines are guides to the eye. Data are displaced by a factor of 1.1 for clarity.}\label{fig:si0p45beta}
\end{figure}

Besides the contribution of $T(Q)$ and background noise, the second term in Eq.~\ref{eq:I(q)forSiO2}, in the form of a stretched exponential decay, describes a combination of contributions from the sensor transfer function~\cite{PhysRevLett.103.194805} and multiple scattering. The existence of multiple scattering is evident based on three observations. Firstly, the sine square term goes to zero periodically, which should make the intensity profile exhibit minima with the same magnitude. However, the measured minima decrease with $q$. The second piece of evidence pointing to the importance of multiple scattering comes from the dynamic data (see later), which displays a $q$-dependent decay rate and exponent that mirrors and anti-mirrors the form of $T(Q)$ (Fig.~\ref{fig:si0p45gamma} and Fig.~\ref{fig:si0p45beta}). This indicates that we are measuring faster dynamics due to multiple scattering where single scattering vanishes. Thirdly, very strongly scattering samples- \emph{i.e.} deliberately multiply scattering samples - show no minima at all.

\begin{figure}
\includegraphics[width=0.9\textwidth]{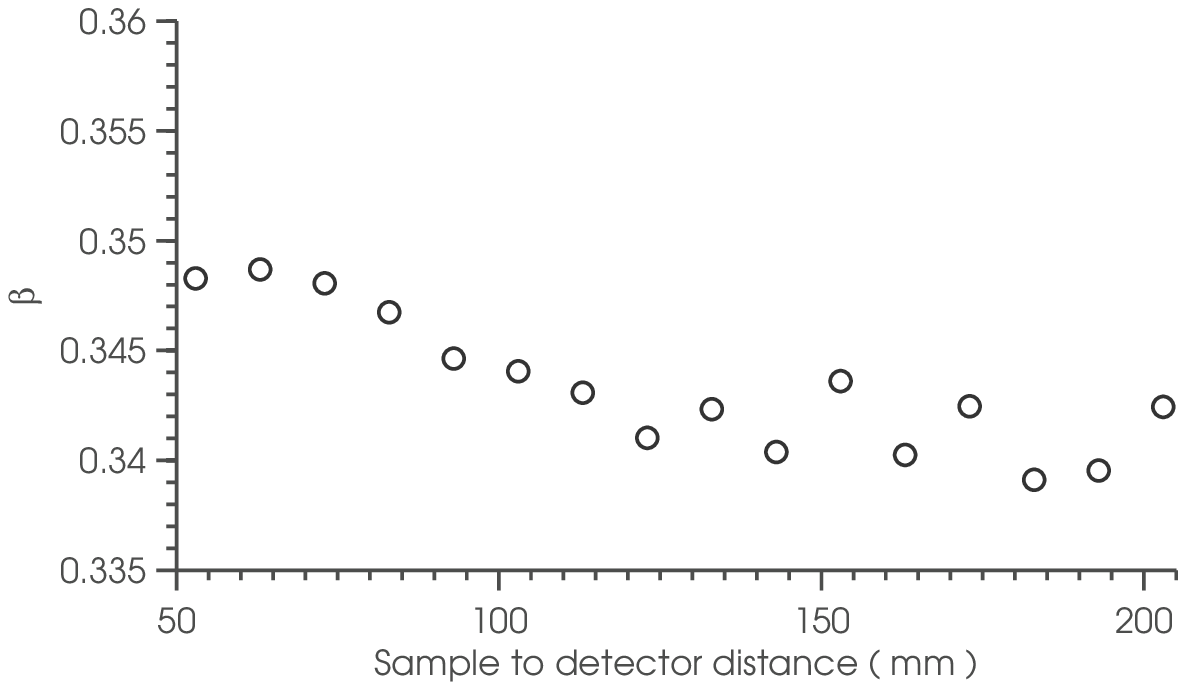}
\caption{The best fit exponent $\beta$ versus sample-to-detector distances varying from 53 mm to 203 mm for silica 0.45 ${\rm{\mu m}}$ suspension.}\label{fig:si0p45intfitpara}
\end{figure}

Hence, we fit the intensity data in two steps. In the first step, we focus on multiple scattering. Note that the single-scattering term vanishes at $q$s satisfying $ \tfrac{1}{2} \left ( - \tfrac{ Q^2 s}{k} + \tfrac{2s}{k \sigma ^2}   + \varphi\right) = n \pi$, where $n=1,2,3...$. As a consequence, we extract the intensity at the $Q$ values corresponding to those minima in $I(Q)$ and fit them with only the multiple scattering term plus the background. The characteristic length $v$ is fixed to $500~\rm{\mu m}$. The amplitude $A_2$ and the exponent $\beta$ were allowed to vary. The resultant multiple scattering term is plotted as the thin solid lines in Fig.~\ref{fig:si0p45intfit}. The best fit exponents $\beta$ are plotted in Fig.~\ref{fig:si0p45intfitpara} versus the sample-to-detector distance $s$. $\beta$ fluctuates around 0.345, giving an empirical stretched-exponential form for the intensity of the multiple scattering. Next, the remaining intensities after subtracting the filled multiple scattering and background are fitted with the transfer function (Eq.~\ref{eq:T(Q)fit}). We use the measured sample-to-detector distance $s$ and try to find one set of $w$, $\xi$ and $\varphi$ that works for all the $s$. The only varying parameter is the overall amplitude. The set $w = 1.8~\rm{\mu m}$, $\xi = 163~\rm{\mu m}$ and $\varphi = 0.026$ yields a good fit for all values of $s$ studied, as shown by the thick solid curves in Fig.~\ref{fig:si0p45intfit}. The value of $w$ provides an estimation of the resolution of our detector, and is close to but somewhat larger than our estimate 1 $\rm{\mu m}$. The small value of $\varphi$ indicates that the x-ray absorption for this silica sample is essentially small. In general, the fitting reproduces the data with few fitting parameters, confirming the theoretical relation between the sample-to-detector distance $s$ and the transfer function $T(q)$, and consequently confirming the feasibility of our experimental setup.

\begin{figure}
\centering
\includegraphics[width=0.8\textwidth]{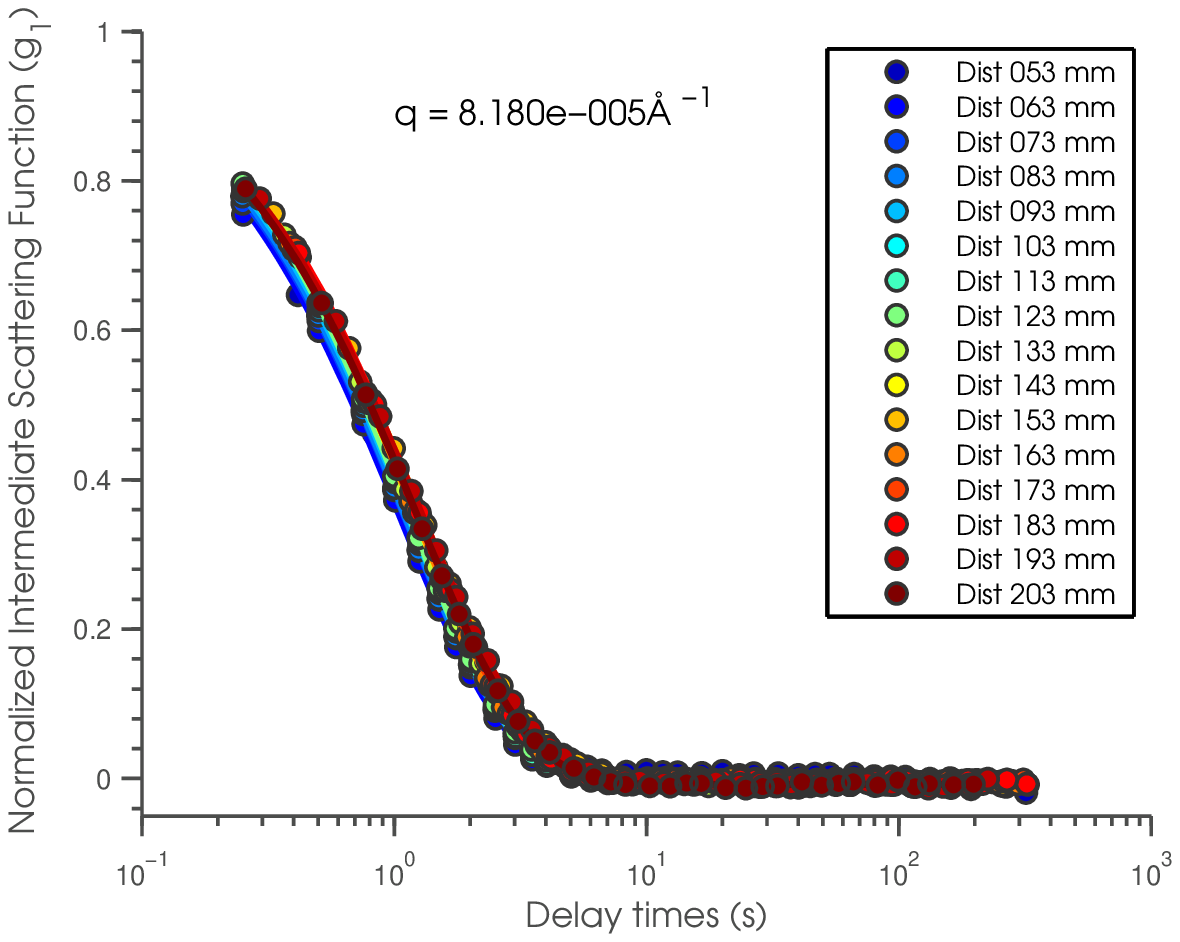}
\caption{Normalized intermediate scattering functions of SiO$_2$ suspension versus delay time $\tau$ measured at different sample-to-detector distances and $q = 8.180 \times 10^{-5}$ \AA.}\label{fig:si0p45g1atdiffz01}
\end{figure}

\begin{figure}
\centering
\includegraphics[width=0.8\textwidth]{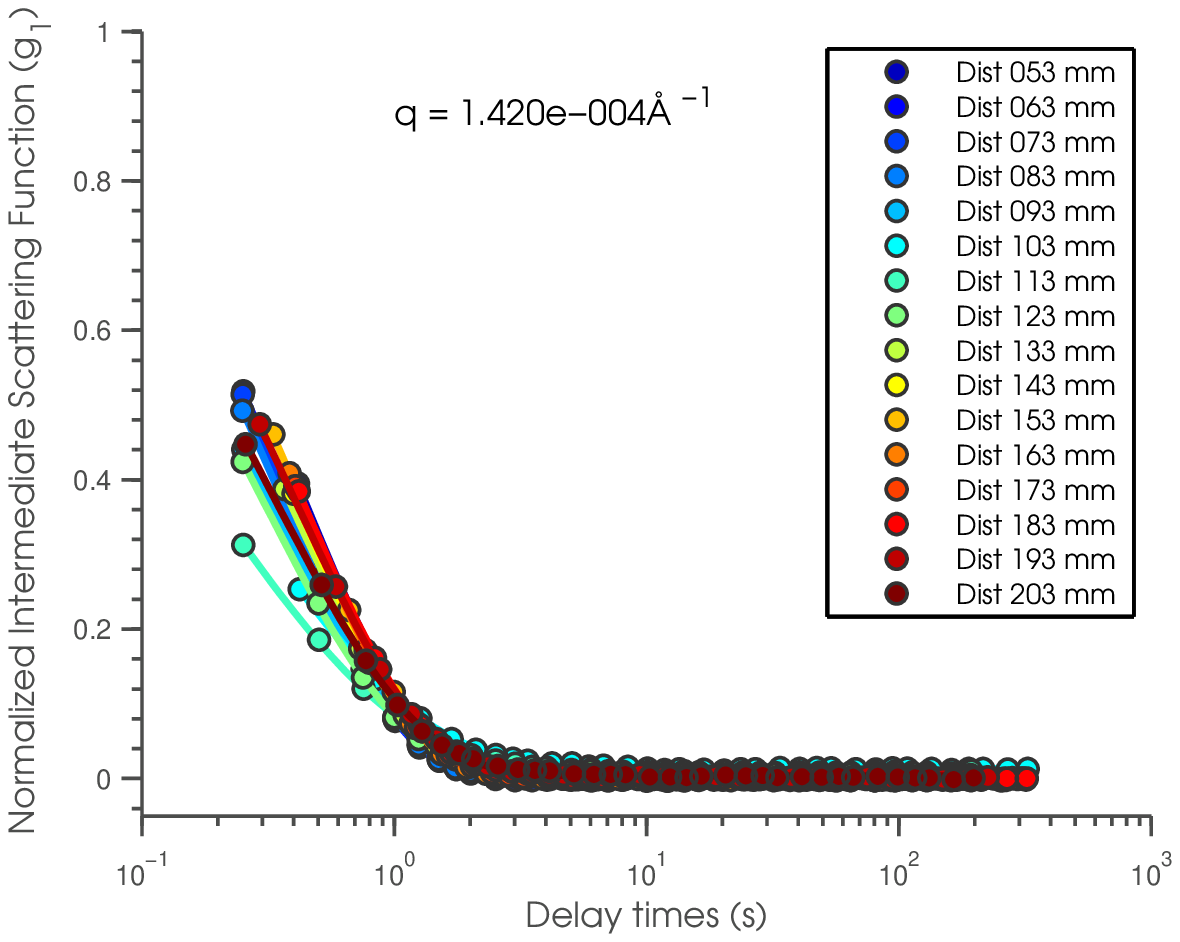}
\caption{Normalized intermediate scattering functions of SiO$_2$ suspension versus delay time $\tau$ measured at different sample-to-detector distances and $q = 1.420 \times 10^{-4}$ \AA$^{-1}$.}\label{fig:si0p45g1atdiffz14}
\end{figure}

With the same principle of XPCS, the fluctuations of near-field speckles should reflect the dynamics of the sample. As shown in Eq.~\ref{eq:g1(q,p)}, the time autocorrelation of the intensity gives rise to $g_1$ instead of $g_2$ in XNFS experiments. Hence, we presented in Fig.~\ref{fig:si0p45g1atdiffz01}, the normalized intermediate scattering function ($g_1(\tau)$) versus delay time ($\tau$) for $\tau$ between 0.4 s and 319 s for different $s$ at $q = 8.180 \times 10^{-5}$ \AA$^{-1}$. The $g_1$s collapse into one curve for different $s$, which agrees with the theoretical prediction that $g_1$ has no $s$-dependence due to the cancelation of $T(q)$. However, for a larger wavevector of $q = 1.420 \times 10^{-4}$ \AA$^{-1}$, the $g_1$s do not overlap for different $s$, as shown in Fig.~\ref{fig:si0p45g1atdiffz14}. To elucidate the reason for this discrepancy and quantify our observations, we have fitted $g_1$ measured at different $s$ and $q$ to a stretched exponential form:
\begin{equation}\label{eq:g1singexp}
    g_1 = e^{-(\Gamma t)^{\alpha}}
\end{equation}

The best-fit relaxation rate ($\Gamma$)(Eq.~\ref{eq:g1singexp}) versus wavevector $q$ is illustrated in Fig.~\ref{fig:si0p45gamma}. The values of $\Gamma$ at successive $s$ are displaced by a factor of 1.1 from the previous $s$ value for clarity. Generally, $\Gamma(q)$ at different $s$ show a $q^2$ behavior, illustrated by the dashed line in Fig.~\ref{fig:si0p45gamma}. However, peaks are observed at the $q$ positions coinciding with the $q$ positions of the dips in the transfer function $T(q)$ (Fig.~\ref{fig:si0p45intfit}).

Away from these multiple scattering peaks, $\Gamma(q)$ increases as $q^2$ versus $q$, which is reasonable for a SiO$_2$ suspension undergoing Brownian motion. Quantitatively, for Brownian motion, we expect $\Gamma(q) = D_m q^2$. As a result, we derive the value of diffusion coefficient $D_m = \Gamma/q^2 \approx 1.167 \times 10^{-12} m^2/s$. According to the first order hydrodynamic interactions $D_m  = D_0 (1 + 1.45\phi )$ \cite{BatchelorJFM}, where Stokes-Einstein diffusion coefficient $ D_0= \frac{{k_B T}}{{6\pi \eta R}} \approx 1.048\times10^{-12} m^2s^{-1}$ with $k_B$ the Boltzmann constant, $T$ the room temperate equal to $293$ K, $\eta$ the dynamic viscosity equal to $1\times 10^3 \rm{kg/(m\cdot s)}$, we obtain $\phi \approx 0.07$, which is reasonable.

At the $q$ positions of the peaks, where the single scattering amplitude goes to minimum because of the zeros of sine$^2$ term in $T(q)$, we hypothesize that we are measuring multiple scattering of the sample. This theory explains why we obtain faster dynamics at those $q$ positions\cite{berne_dynamic_2000}. Fig.~\ref{fig:si0p45beta} shows the corresponding best fit exponent $\alpha$, which exhibits similar fluctuation patterns as $T(q)$ and supports our hypothesis. Underlying this hypothesis is the idea that the rapid variations of $T(q)$ versus $q$ may be associated with single scattering, whereas the intensity of multiple scattering likely shows a relatively smooth $q$-dependence. Accordingly,  if, for a particular set of data, the scattering minima due to $T(q)$ are indistinct (do not send the scattering intensity to zero) then it follows that the XNFS data set in question suffers from multiple scattering. Hence, to calculate $g_1$, we have to pick $q$ smaller than the first dip of $T(q)$ so that the measured coherent scattering is reliable.

\begin{figure}
\centering
\includegraphics[width=0.9\textwidth]{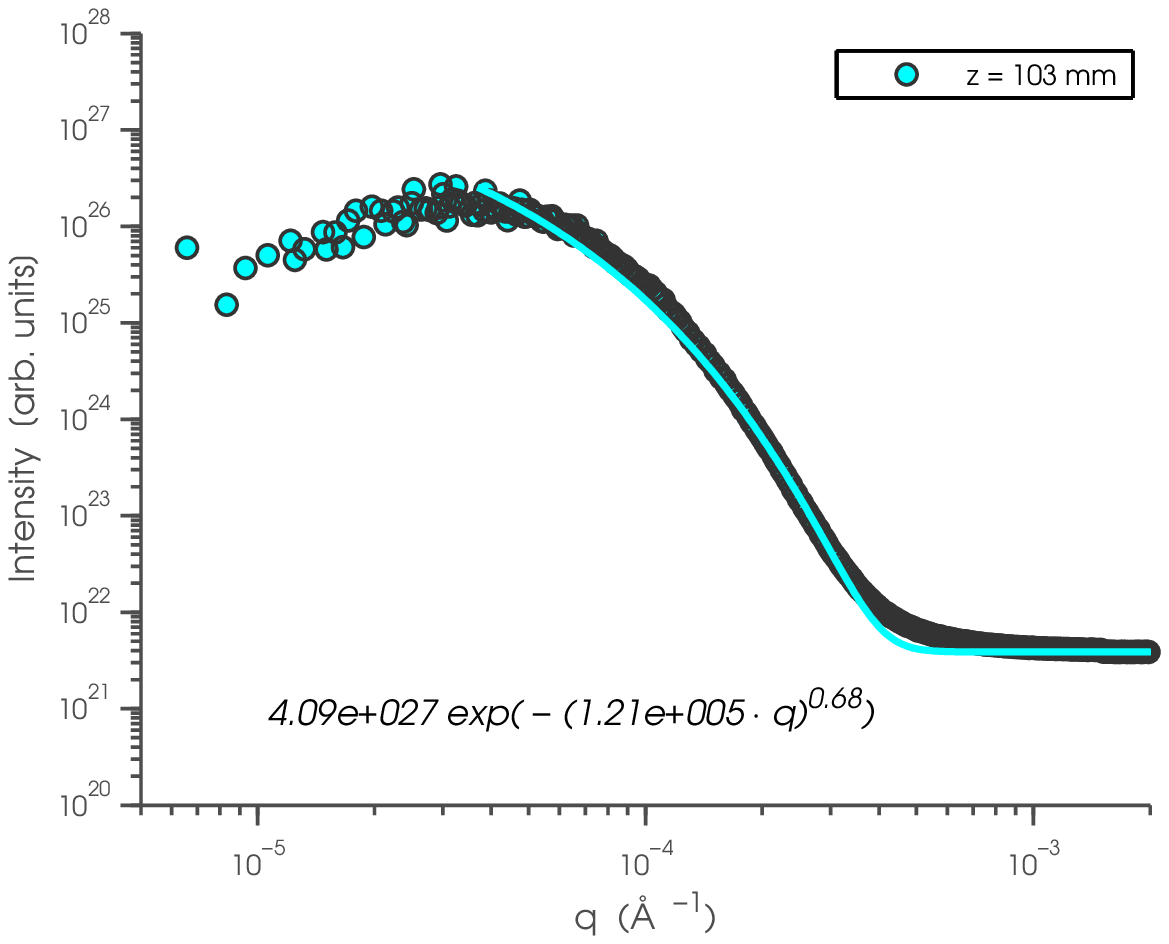}
\caption{Intensities of 3 mm thickness Gillette shaving foam at sample-to-detector distance $s = 103$ mm. The symbols are the data. And the lines are the fittings based on a stretched exponential decay $I(q) = C \exp(-(A \cdot q)^{\beta})$.}.\label{fig:foamintfit}
\end{figure}

\begin{figure}
\centering
\includegraphics[width=0.9\textwidth]{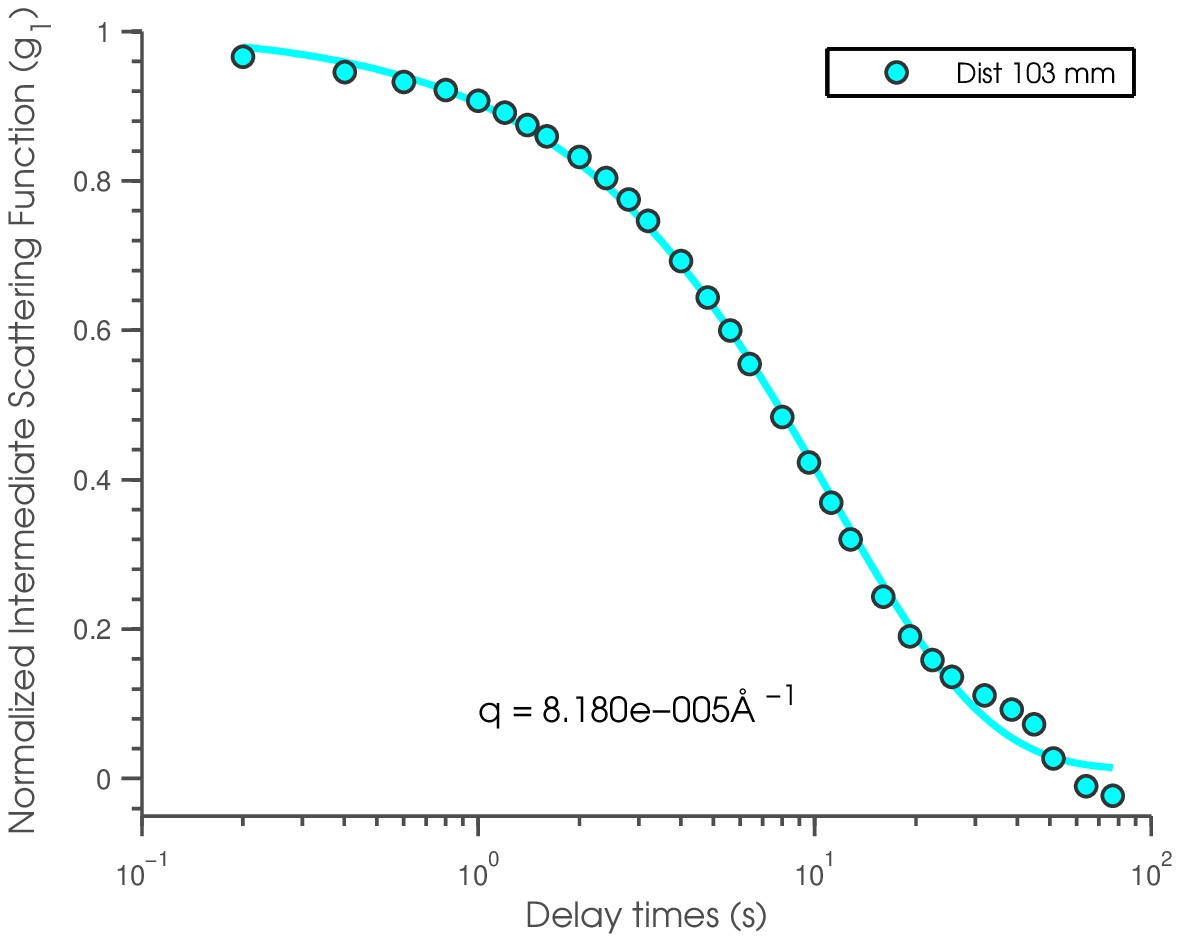}
\caption{Normalized intermediate scattering functions of 3 mm thickness Gillette shaving foam at sample-to-detector distance $s = 103$ mm and $q=8.18 \times 10^{-5}$ \AA. The symbols are the data. And the lines are the fittings based on a stretched exponential decay (Eq.~\ref{eq:g1singexp}).}.\label{fig:foamg1fit}
\end{figure}


To further test this idea, we carried out measurements on a sample that could be expected to show very strong scattering and therefore strong multiple scattering, namely a 3 mm-thick sample of Gillette Foamy shaving foam, which is know to consist of a dense foam of micron-sized air bubbles in aqueous liquid. Fig.~\ref{fig:foamintfit} shows the scattering intensity from such a sample as a function of $q$, obtained using the XNFS prescription. However, in contrast to the more-weakly scattering silica spheres, discussed above, evidently, in this case there are not the oscillations in intensity that are expected for XNFS, i.e. there is no evidence that the XNFS $|T(q)|^2$ is displayed in these data. We infer that this is indeed the result of multiple scattering and that the x-ray scattering from 3 mm-thick foamy is completely in the multiple scattering regime. This implies that $|T(q)|^2$ is a signature of single scattering. We can also calculate $g_1$ for foamy according to the XNFS prescription. This is shown in Fig.~\ref{fig:foamg1fit}. The dynamics is pretty slow, on the order of 0.1 $s^{-1}$.

These results point to another difficulty with the XNFS method (which is common to ultra-small-angle x-ray scattering methods in general) namely that multiple scattering must be carefully considered and if possible eliminated. In the case of foamy, a sufficiently thin sample (much thinner than 3 mm) would have eventually had reached the single scattering regime. Interestingly, in the case of XNFS, in contrast to more traditional USAXS methods, the existence or not of multiple scattering may be straightforwardly and immediately recognized from the intensity profile i.e. $|T(q)|^2$ as we discussed previously.

\section{Polystyrene 4 $\mathbf{\mu m}$ suspension}

\begin{figure}
\centering
\includegraphics[width=1\textwidth]{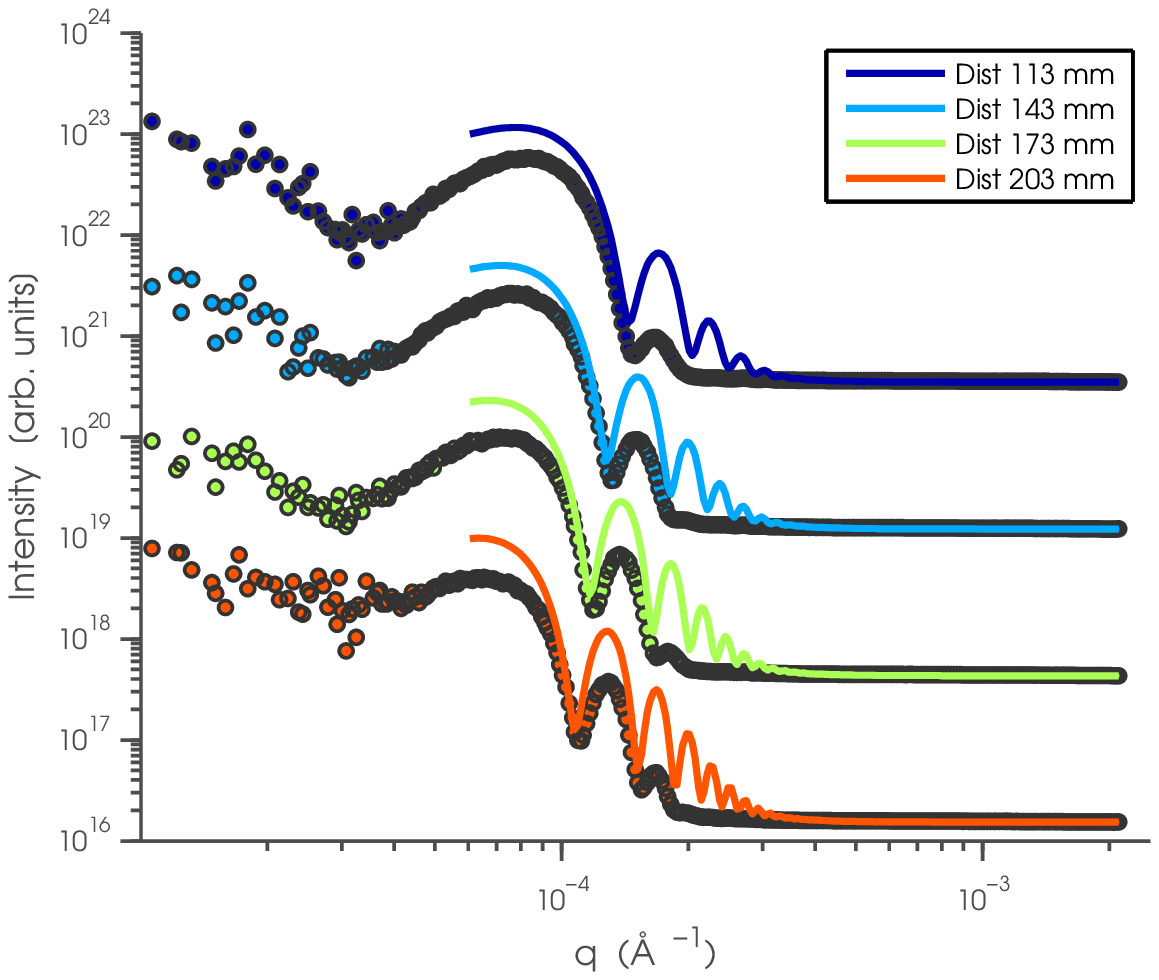}
\caption{Intensities of Polystyrene 4 $\rm{\mu m}$ suspension at different sample-to-detector distances. The disks with different colors are the data measured at different sample to detector distance. The solid lines are theoretical plots of the transfer function $T(q)$ of the silica suspension for comparison. The data are displaced for clearance.}\label{fig:ps04int}
\end{figure}

In this section, we present the XNFS data obtained from a colloidal suspension of polystyrene particles of a diameter of 4 $\rm{\mu m}$. This sample is not as stable as the last sample, since particles with 4 $\rm{\mu m}$ undergo sedimentation. The static structure factor peak of the polystyrene suspension of this size lies within the $q$-range accessible by our XNFS setup. Hence, we expect to observe more complicated intensity profiles with the contributions from both structure of the suspension and the transfer function. Illustrated in Fig.~\ref{fig:ps04int} are the scattering intensities (symbols) plotted versus $q$ obtained by azimuthally averaging the fourier transformed scattering images averaged over 1000 frames for sample-to-detector distances $s = 113$ mm, $143$ mm, $173$ mm and $203$ mm (from top to bottom). The corresponding transfer functions ($T(q)$) obtained by fitting data of silica sample measured at the same $s$ are plotted in solid lines with the same colors for easy comparisons for the peak positions of $T_{talb}$. The intensity data deviate from the lines. Firstly, the peak positions of the data matches the ones of $T(q)$. There might be one extra peak located at $q$ around $10^{-4}$ \AA$^{-1}$ for all $s$, which is coming from the static structure factor peak. In addition, the peaks of this sample are less sharp than the ones of silica sample. We attribute this to the fact that the scattering contrast of polystyrene is less than that of silica, leading to weaker scattering for the polystyrene sample, even though the PS particle are larger. 

\begin{figure}
\centering
\includegraphics[width=0.95\textwidth]{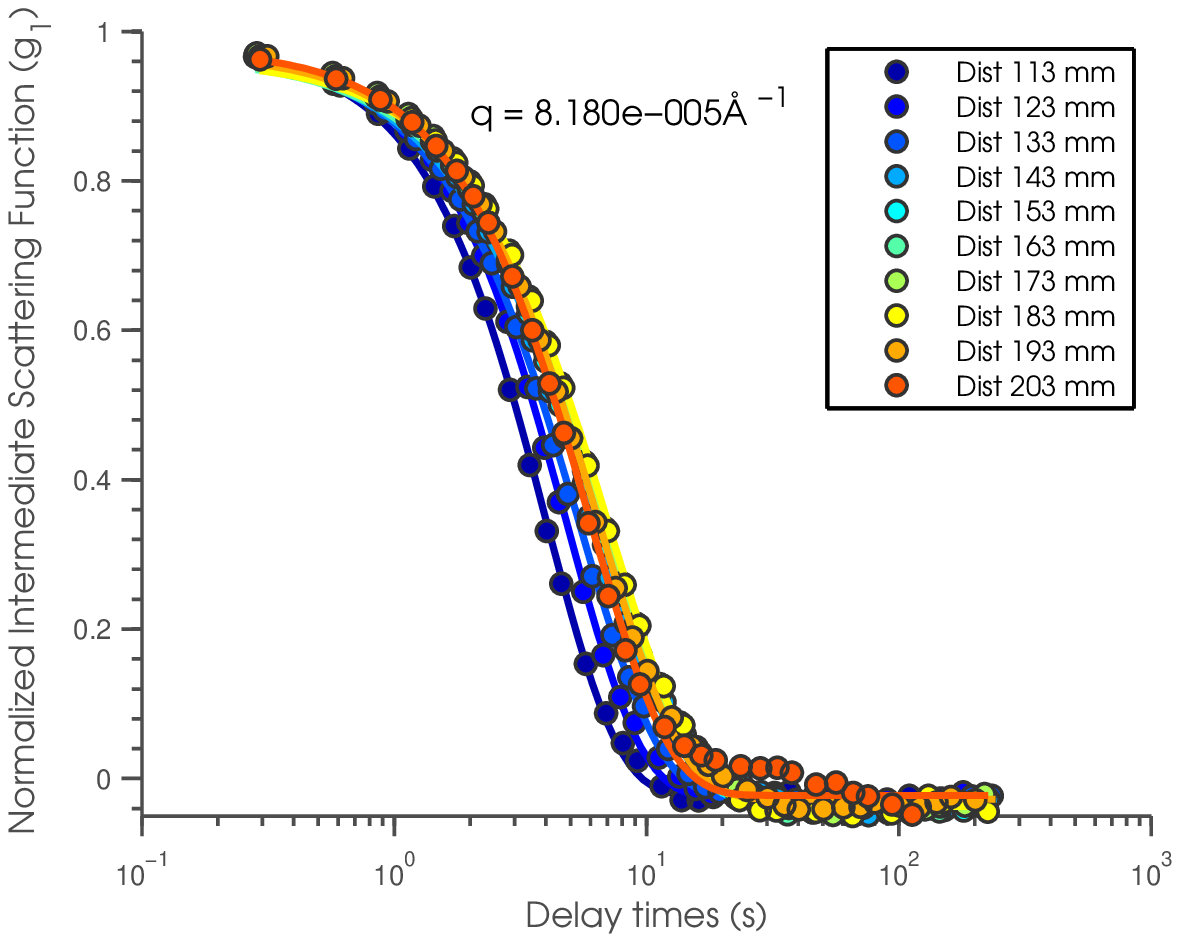}
\caption{Normalized Autocorrelation Functions of Polystyrene 4 $\rm{\mu m}$ suspension measured at different sample-to-detector distances.}\label{fig:ps04g1atdiffz}
\end{figure}

Illustrated in Fig.~\ref{fig:ps04g1atdiffz} are the normalized intermediate scattering functions ($g_1$) at $q = 8.18 \times 10^{-5}$ \AA$^{-1}$ for delay time from $0.02D_0q^2$ to $20D_0q^2$ seconds and sample-to-detector distances $s$ from $203$ mm to $113$ mm with an interval of $10$ mm. The $g_1$s for the polystyrene suspension do not totally overlap for different $s$ at this $q$ position, but decays slightly faster when the detector moves closer to the sample stage. We reason that this is the result of the sedimentation of the polystyrene particles, which leads to denser sample with faster dynamics.

\begin{figure}
  \includegraphics[width=1\textwidth]{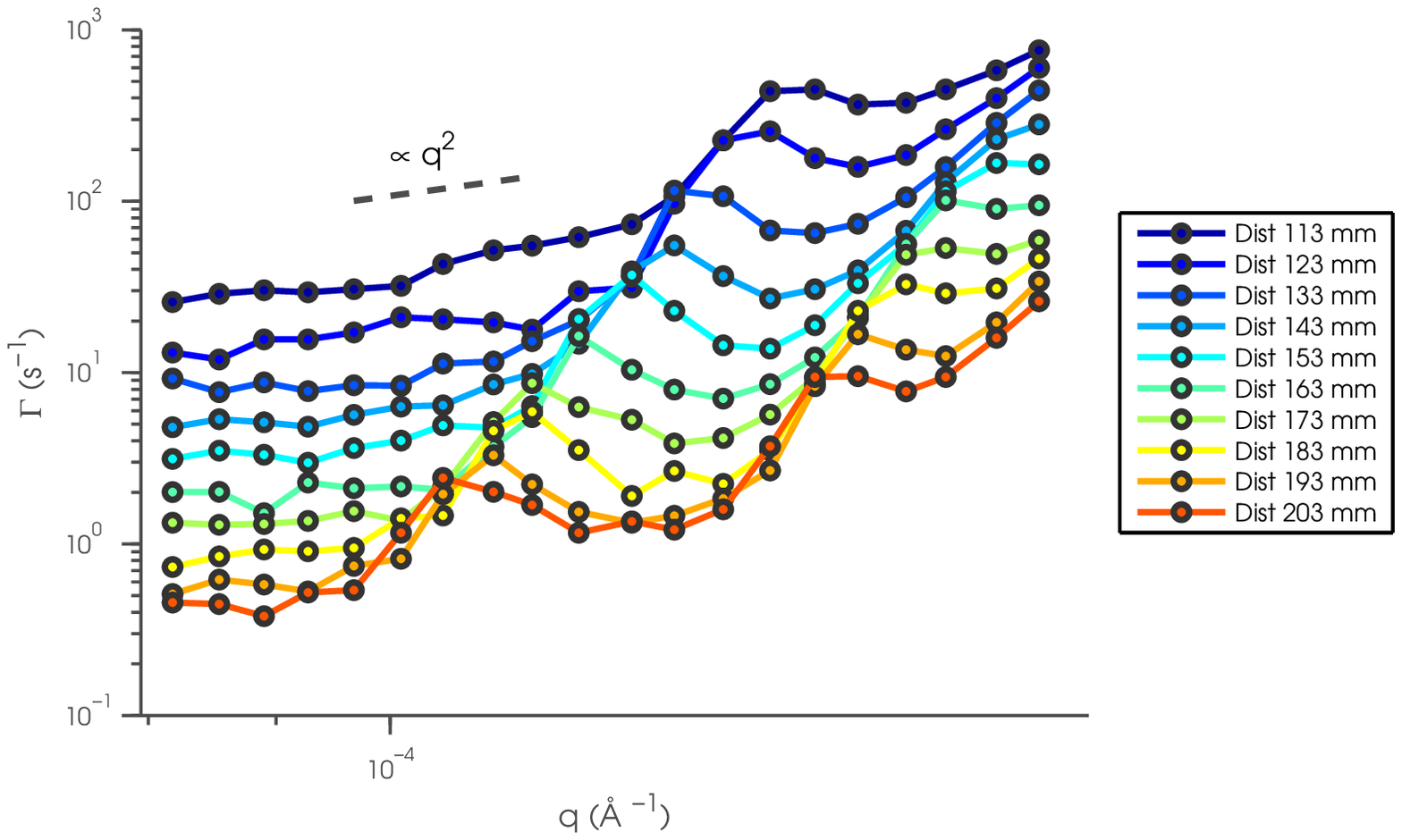}\\
  \caption{The best-fit decay rate $\Gamma$ versus wavevector $q$ for polystyrene of diameter 4 $\rm{\mu m}$ measured at different $s$. Points are the fitting results and the lines are guides to the eye. Data are displaced for clarity.}\label{fig:ps04gammavsq}
\end{figure}

Following the same procedure as for the silica sample, the best-fit relaxation rates ($\Gamma$), obtained by fitting the one of the 100-frames $g_1$ with single exponential form (Eq.~\ref{eq:g1singexp}), are plotted versus wavevector $q$ for different $s$ in Fig.~\ref{fig:ps04gammavsq}. The values of $\Gamma$ at different $s$ are displaced for clarity. Similarly, peaks that correspond to dips of transfer function are observed, confirming our conclusion about the measurement of multiple scattering at the minimums of transfer function. In this case, the peaks are more visible than the ones observed in silica sample, indicating stronger multiple scattering in this sample with bigger polystyrene particles.

\section{Future work and Conclusion}

In conclusion, we have presented the implementation of the new coherent x-ray technique - X-ray Near Field Speckle as well as its applications and limitations. Clearly, XNFS is capable of obtaining ultra small angle x-ray scattering and x-ray photon correlation spectroscopy with its simple setup and direct relationship to the density correlation function. It effectively extends to wavevectors an order of magnitude smaller than the wavevector range covered by conventional SAXS and XPCS, and enables us to explore the static and dynamic structures of micrometer-sized samples. We believe this technique will be valuable for optically dense and turbid samples which induce strong multiple scattering optically.

Technically, XNFS is not difficult to realize. Speckle pattern is produced simply by letting both scattered beam and transmitted beam impinge onto the detector. It does not require the spatial filtering as did in XPCS, which allows us using the whole source beam and in turn greatly enhance the speckle contrast. As a consequence, it does not require laborious alignments. All the efforts were devoted to the design of the detector. High numerical aperture objective was employed to produce high spacial resolution and efficient light collection. The measurements gives convincing results, which proves the feasibility of this setup. Improvements could be made on several aspects. One is to utilize thinner scintillator, which will give rise to less spherical aberrations. A faster CCD camera will for sure improve the probing range of the dynamics of this technique.

One key difficulty of this technique is due to the transfer function $T(q)$. It entangles with S(q). it is straightforward to characterize the structure factor peaks and dips located smaller than the $q$ position of the first dip of $T(q)$. However, this would make the reliable $q$ range very small. If XNFS is to realize its full potential, it will be necessary to figure out an effective way to deconvolve the static structure factor from the transfer function in the future. One possible way is to use as small as possible sample-to-detector distance though with a cost in scattering contrast. Strong absorption samples might not be affected by this factor due to the phase factor induced in the sine term of $T(q)$(Eq.~\ref{eq:T(Q)fit}). Another possible improvement might be made by measuring a control sample with exactly the same material but uniform $S(q)$ in the accessible $q$ window, then dividing the intensity profile of the interested sample and the control sample. Another important issue is multiple scattering. Evidence of the existence of multiple scattering comes from the intensity profile and sample-to-detector dependent decay rate. Making the sample as thin as possible should solve this problem.

Finally, the limited dynamic range of the CCD (12 bits) means that weakly scattering samples cannot be studied, because they give a scattered intensity that is less than $1/4000$ of the direct beam intensity and so their scattering contribution cannot be reached. Using a 16, 18 or 20 bit CCD would extent the range of possible XNFS sample correspondingly.

\section{Acknowledgements}

We thank T. Chiba, A. Mack, E. R.  Dufresne, R. L. Leheny, C. O'Hern, S. Sanis, M. Spannuth, and J. Wettlaufer for discussions and the NSF for support via DMR 0453856. The APS is supported by the DOE.

\bibliography{./simon}

\end{document}